\documentclass[11pt]{article}

\usepackage{pstricks}
\usepackage{pst-plot}
\usepackage{pst-node}
\usepackage{amsmath}
\usepackage{pst-grad}
\usepackage{amssymb}
\usepackage{xspace}
\usepackage{pst-math}
\usepackage{pst-xkey}
\usepackage{pst-coil}
\usepackage{pst-3dplot}
\usepackage[T1]{fontenc}
\usepackage{ae}
\usepackage[ansinew]{inputenc}

\newrgbcolor{darkred}{0.8 0 0}
\newrgbcolor{darkerred}{0.7 0 0}
\newrgbcolor{darkestred}{0.5 0 0}
\newrgbcolor{darkgreen}{0 0.5 0.5}
\newrgbcolor{darkergreen}{0 0.6 0}
\newrgbcolor{darkestgreen}{0 0.5 0}
\newrgbcolor{darkblue}{0 0 0.6}
\newrgbcolor{darkestblue}{0 0 0.5}
\newrgbcolor{redgreen}{0.8 0.8 0}
\newrgbcolor{bluegreen}{0 0.4 0.7}
\newrgbcolor{bluewhite}{0 0.9 1}
\newrgbcolor{lightgreen}{0.6 1 0.6}
\newrgbcolor{lightred}{1 0.8 0.8}
\newrgbcolor{lightyred}{1 0.4 0.4}
\newrgbcolor{lightyblue}{0.5 0.5 1}
\newrgbcolor{lightblue}{0.8 1 1}
\newrgbcolor{lilas}{0.5 0.3 0.7}
\newrgbcolor{lilaswhite}{1 0.8 1}
\newrgbcolor{redy}{1 0.6 0.6}
\newrgbcolor{bluish}{0.6 0.6 1}
\newrgbcolor{sand}{0.9 0.9 0.7}
\newrgbcolor{brown}{0.4 0.4 0}
\newrgbcolor{blue1}{0 0.35 0.65}
\newrgbcolor{blue2}{0 0.4 0.6}
\newrgbcolor{blue3}{0 0.45 0.55}
\newrgbcolor{blue4}{0 0.5 0.5}
\newrgbcolor{blue5}{0 0.55 0.45}
\newrgbcolor{blue6}{0 0.6 0.4}
\newrgbcolor{blue7}{0 0.65 0.35}

\newcommand{\R }{\mathbb R}
\newcommand{\N }{\mathbb N}
\newcommand{\Z }{\mathbb Z}
\newcommand{\Q }{\mathbb Q}
\newcommand{\C }{\mathbb C}
\newcommand{\sen }{\ \! {\rm sen}\ \! }
\newcommand{\tg }{\ \! {\rm tg}\ \! }
\newcommand{\cosec }{\ \! {\rm cosec}\ \! }
\newcommand{\cotg }{\ \! {\rm cotg}\ \! }
\newcommand{\dx }{\ \! dx}
\newcommand{\e }{\ \! e}
\newcommand{\senh }{\ \! {\rm senh}\ \! }
\newcommand{\tgh }{\ \! {\rm tgh}\ \! }
\newcommand{\cotgh }{\ \! {\rm cotgh}\ \! }
\newcommand{\sech }{\ \! {\rm sech}\ \! }
\newcommand{\cosech }{\ \! {\rm cosech}\ \! }
\newcommand{\arcsen }{\ \! {\rm arcsen}\ \! }
\newcommand{\arctg }{\ \! {\rm arctg}\ \! }
\newcommand{\arccotg }{\ \! {\rm arccotg}\ \! }
\newcommand{\arcsenh }{\ \! {\rm arcsenh}\ \! }
\newcommand{\arccosh }{\ \! {\rm arccosh}\ \! }
\newcommand{\arctgh }{\ \! {\rm arctgh}\ \! }
\newcommand{\arccosech }{\ \! {\rm arccosech}\ \! }
\newcommand{\arcsech }{\ \! {\rm arcsech}\ \! }
\newcommand{\arcsec }{\ \! {\rm arcsec}\ \! }
\newcommand{\arccosec }{\ \! {\rm arccosec}\ \! }
\renewcommand{\arctgh }{\ \! {\rm arctgh}\ \! }
\newcommand{\arccotgh }{\ \! {\rm arccotgh}\ \! }
\newcommand{\nega }{\neg \ }
\newcommand{\eq }{\Leftrightarrow }

\begin{document}

\title{A Map of the Brazilian Stock Market}

\author{Leonidas Sandoval Junior \\ \\ Insper, Instituto de Ensino e Pesquisa}

\maketitle

\begin{abstract}
We use the correlation matrix of stocks returns in order to create maps of the São Paulo Stock Exchange (BM\&F-Bovespa), Brazil's main stock exchange. The data reffer to the year 2010, and the correlations between stock returns lead to the construction of a minimum spanning tree and of asset graphs with a variety of threshold values. The results are analised using techniques of network theory.
\end{abstract}

\section{Introduction}

The BM\&F-Bovespa (Bolsa de Valores, Mercadorias e Futuros de São Paulo) is the major stock exchange in Brazil, with a market capitalization of US\$ 927 million in 2010. It is also a good representative of a stock exchange of an emerging market, and has received increasing attention by international investors in the past years.

The aim of this article is to present a map, actually two maps, of the BM\&F-Bovespa in 2010. In order to do so, I shall use some techniques taken from Random Matrix Theory \cite{rmt4}, first developed for the use in nuclear physics and then used in many areas, including finance (see \cite{leocorr} for a comprehensive list of contributions). There are many studies of networks built from data of financial markets around the world, mainly based on the New York Stock Exchange \cite{h1}-\cite{h37}, but also using data from Nasdaq \cite{h37}, the London Stock Exchange \cite{h29} \cite{h14}, the Tokyo Stock Exchange \cite{h29} \cite{h42}, the Hong Kong Stock Exchange \cite{h29}, the National Stock Exchange of India \cite{h16}, the Global financial market \cite{h13} \cite{h29} \cite{h19}-\cite{h27}, the USA Commodity market \cite{h30}, the foreign currency market \cite{h31}-\cite{h39}, and the world trade market \cite{h33}-\cite{h36}. Until the present date, to the author's knowledge, no work has been done in using the stocks of BM\&F-Bovespa as a source of data for developing networks.

Both maps are done using the time series of the 190 stocks of BM\&F-Bovespa that were negotiated every day the stock exchange was open (a list is given in Appendix A), so that those stocks are all very liquid. From the time series of daily prices, I obtained the series of log-returns, given by
\begin{equation}
S_t=\ln (P_t)-\ln (P_{t-1})\approx \frac{P_t-P_{t-1}}{P_t}\ ,
\end{equation}
where $P_t$ is the price of a stock at day $t$ and $P_{t-1}$ is the price of the same stock at day $t-1$. The correlation matrix between all log-returns was then calculated using the data obtained for the whole year of 2010.

The time series of stocks prices encode an enormous amount of information about the way they relate to each other, and only part of that information is captured by the correlation matrix of their log-returns. This information also presents a good amount of noise, and should be filtered whenever that is possible. In this work, I employ some threshold values based on simulations of randomized data based on the time series that is under study in order to eliminate some of that noise. The shuffled data is obtained by reordering in a random way every time series of every single log-return. This generates time series that have the exact probability distribution as the original data, but with the connections between each log-return made completely random. The correlation matrix obtained from the randomized data is then compared with the correlation matrix of the original data. All connection values that are of comparable to the ones of randomized data are then eliminated or marked as possibly random.

There are many measures of correlation between elements of time series, the most popular being the Pearson correlation coefficent. The drawback of this correlation measure is that it only detects linear relationships between two variables. Two different measures that can detect nonlinear relations are the Spearman and the Kendall tau rank correlations, which measure the extent to which the variation of one variable affects other variable, withouth that relation being necessarily linear. In this work, I chose Spearman's rank correlation, for it is fairly fast to calculate and it is better at measuring nonlinear relations.

The correlation matrix may then be used to create a distance matrix, where distance is a measure of how uncorrelated two log-return series are from one another. There are also many ways to build a distance measure from the correlations between data. In this work, I shall use a different metric from \cite{h1}, which is a nonlinear mapping of the Pearson correlation coefficients between stock returns. The metric to be considered here differs from the aforementioned metric because it is a linear realization of the Spearman rank correlation coefficient between the indices that are being studied:
\begin{eqnarray}
\label{metric2}
 & & d_{ij}=1-c_{ij}\ ,
\end{eqnarray}
where $c_{ij}$ are elements of the correlation matrix calculated using Sperman's rank correlation. This distance goes from the minimum value 0 (correlation 1) to the maximum value 2 (correlation -1).

Using the distance matrix, I shall build two maps (networks) based on the stocks of BM\&F-Bovespa. The first map shall be based on a Minimum Spanning Tree (MST), which is a graph where each stock is a node (vertex), connected to one or more nodes. An MST is a planar graph with no intersections in which all nodes are connected and the sum of distances is minimum. The number of connections is the same as the total number of nodes of the network, minus one. This type of representation is very useful for visualizing connections between stocks, but it often oversimplifies the original data. Another representation is called an {\sl asset graph}, which may be built by establishing a threshold under which correlations are considered, eliminating all other correlations above the said threshold. This also makes the number of original connections drop, simplifying the information given by the original correlation matrix. A three-dimensional map of stocks may be built in such a way that the distances portraied in it are the best approximation to the real distances, and the connections obtained by establishing a threshold are then drawn on such representation.

The maps are examples of two networks that can be obtained from the same original data, each with its advantages and drawbacks. Using the two maps, one may then be able to identify which stocks are more dependent on each, and also which ones are more connected to others, what may be useful when building portfolios that minimize risk through diversification \cite{Elton}. There are measures of how central each node of a network is, establishing its overall importance in the web of nodes, and also ways to visualize the overall distribution of those measures. As we shall see, the maps help verify that stocks that belong to the same types of companies tend to aglommerate in the same clusters, and that stocks with weak correlations often tend to connect at random in those networks.

The MST is built in section 2, its centrality measures are shown in section 3, and their cumulative distribution functions are studied in section 4. The assets graphs are built in section 5, and the centrality of one of them are studied in section 6. Section 7 discusses the k-shell decomposition of one of the asset graphs, and section 8 presents a conclusion and general discussion of results.

\section{Minimum spanning trees}

Minimum spanning trees are networks of nodes that are all connected by at least one edge so that the sum of the edges is minimum, and which present no loops. This kind of tree is particularly useful for representing complex networks, filtering the information about the correlations between all nodes and presenting it in a planar graph.

As discussed in the introduction, I shall employ simulations of randomized data in order to establish a distance threshold above which correlations are seen as possibly of random nature. The result of 1000 simulations of randomized data is a lower threshold $0.69\pm 0.02$, so distances above this value are represented as dashed lines in the minimum spanning tree diagrams.

The minimum spanning tree formed by the stocks of the BM\&F-Bovespa which were negotiated every day the stock exchange functioned during 2010 is represented in Figure 1. It has four main hubs: those stocks are VALE3 and VALE5, belonging to Vale, the major mining industry in Brazil and one of the largest in the world, BBDC4, stock from Bradesco, one of the major banks in Brazil, and BRAP4, which is a branch of Bank Bradesco responsible for its participations in other companies. Of local importance are GFSA3, stocks from GAFISA, and PDGR3, stocks from PDG Realty, both construction and materials companies. The stocks are color-coded by main economic sectors, as shown in Figure 2.



\vskip 1.4 cm

\begin{center}
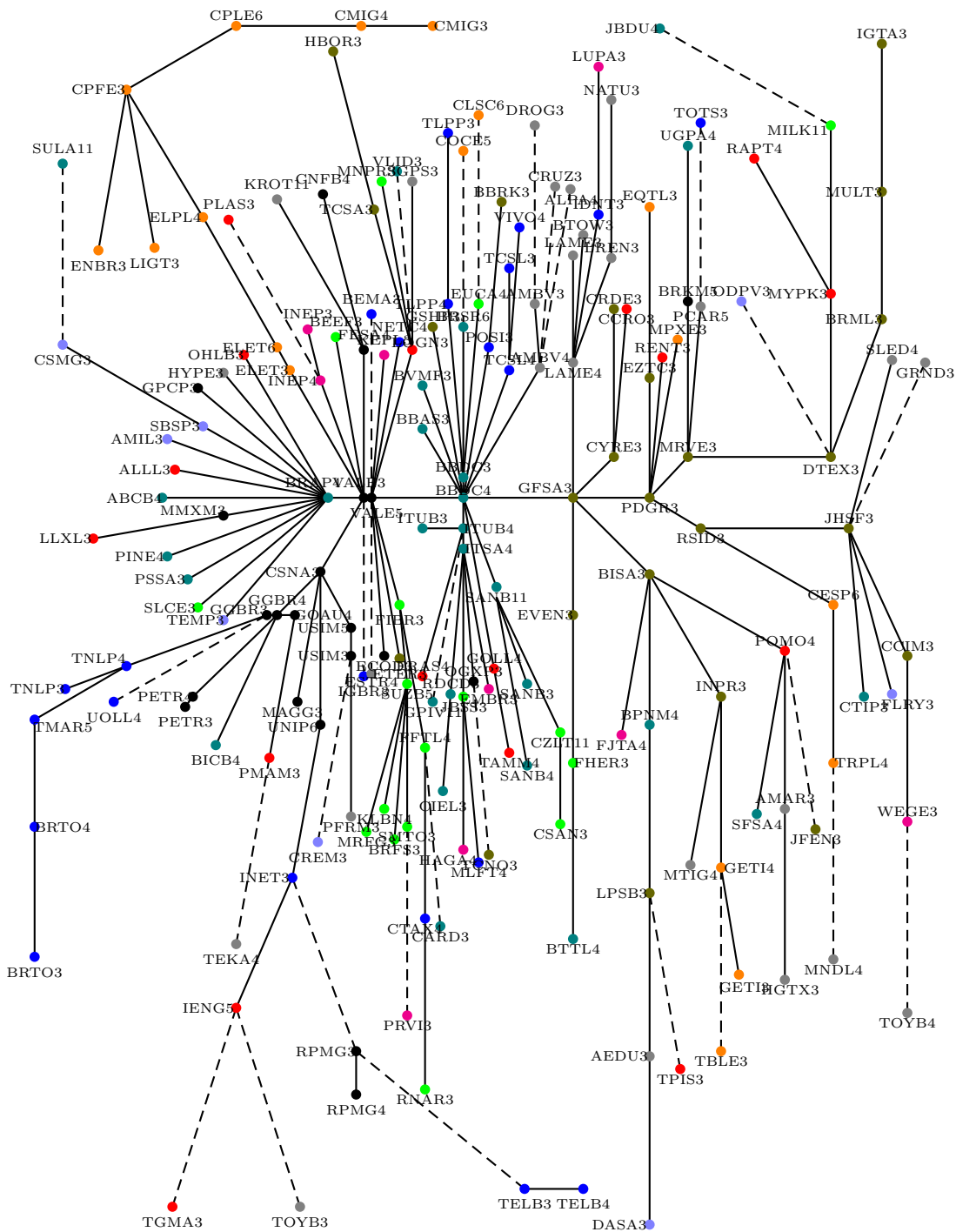
{\bf Figure 1.} Minimum spanning tree for 2010.
\end{center}

\vskip 0.4 cm 

%

\begin{center}
{\bf Figure 2.} Color coding of stocks in Figure 1 according to economic sectors.
\end{center}

\vskip 0.3 cm

The network formed by the stocks of the BM\&F-Bovespa which were negotiated every day the stock exchange functioned during 2010 have six main hubs, which are displayed in Figure 3. The stocks are VALE3 and VALE5, belonging to Vale, the major mining industry in Brazil and one of the largest in the world, BBDC4, stock from Bradesco, one of the major banks in Brazil, BRAP4, which is a branch of Bank Bradesco responsible for its participations in other companies, GFSA3, stocks from GAFISA, and PDGR3, stocks from PDG Realty, both construction and materials companies.

\begin{pspicture}(-5,-0.4)(1,1.5)
\psset{xunit=8,yunit=8} \scriptsize
\psline(0,0)(-0.03,0) 
\psline(-0.03,0)(-0.17,0) 
\psline(0,0)(0.36,0) 
\psline(0.36,0)(0.79,0) 
\psline(0.79,0)(1.09,0) 
\psdot[linecolor=black,linewidth=1.2pt](0,0) 
\psdot[linecolor=black,linewidth=1.2pt](-0.03,0) 
\psdot[linecolor=darkgreen,linewidth=1.2pt](-0.17,0) 
\psdot[linecolor=darkgreen,linewidth=1.2pt](0.36,0) 
\psdot[linecolor=brown,linewidth=1.2pt](0.79,0) 
\psdot[linecolor=brown,linewidth=1.2pt](1.09,0) 
\rput(0.02,-0.04){VALE5}
\rput(-0.05,0.04){VALE3}
\rput(-0.28,0){BRAP4}
\rput(0.36,0.04){BBDC4}
\rput(0.79,0.04){GFSA3}
\rput(1.19,0){PDGR3}
\end{pspicture}

\begin{center}
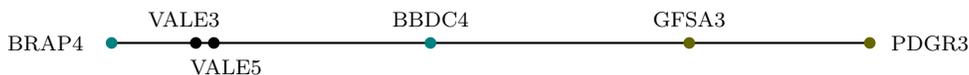
{\bf Figure 3.} Main axis of the minimum spanning tree for 2010.
\end{center}

For reasons of clarity, I shall divide the minimum spanning tree into five diagrams, or clusters, each one constructed around one of the five main hubs. The first cluster we shall study in more detail, Figure 4, is the one formed around BRAP4, Bradespar, which is a company created when Bradesco (banking) was dismembered. It manages the participations of Bradesco in other, non-financial companies, particularly CPFL and Vale. 

\begin{pspicture}(-11.5,-3.9)(1,4)
\psset{xunit=5,yunit=5} \scriptsize
\psline[linecolor=blue](0,0)(0.14,0) 
\psline(0,0)(-0.56,0) 
\psline(-0.56,0)(-1.20,0) 
\psline[linestyle=dashed](-1.20,0)(-1.91,0) 
\psline(0,0)(-0.36,0.21) 
\psline(-0.36,0.21)(-0.80,0.46) 
\psline(0,0)(-0.34,0.58) 
\psline(0,0)(0,0.64) 
\psline(0,0)(0.34,0.58) 
\psline(0,0)(0.54,0.32) 
\psline(0,0)(-0.59,-0.27) 
\psline(0,0)(-0.39,-0.47) 
\psline(0,0)(-0.17,-0.65) 
\psline(0,0)(0.12,-0.66) 
\psline(0,0)(0.37,-0.52) 
\psline(0,0)(0.56,-0.32) 
\psdot[linecolor=darkgreen,linewidth=2.2pt](0,0) 
\psdot[linecolor=black,linewidth=2.2pt](0.14,0) 
\psdot[linecolor=lightyblue,linewidth=2.2pt](-0.56,0) 
\psdot[linecolor=lightyblue,linewidth=2.2pt](-1.20,0) 
\psdot[linecolor=darkgreen,linewidth=2.2pt](-1.91,0) 
\psdot[linecolor=black,linewidth=2.2pt](-0.36,0.21) 
\psdot[linecolor=red,linewidth=2.2pt](-0.80,0.46) 
\psdot[linecolor=darkgreen,linewidth=2.2pt](-0.34,0.58) 
\psdot[linecolor=darkgreen,linewidth=2.2pt](0,0.64) 
\psdot[linecolor=green,linewidth=2.2pt](0.34,0.58) 
\psdot[linecolor=lightyblue,linewidth=2.2pt](0.54,0.32) 
\psdot[linecolor=darkgreen,linewidth=2.2pt](-0.59,-0.27) 
\psdot[linecolor=red,linewidth=2.2pt](-0.39,-0.47) 
\psdot[linecolor=lightyblue,linewidth=2.2pt](-0.17,-0.65) 
\psdot[linecolor=black,linewidth=2.2pt](0.12,-0.66) 
\psdot[linecolor=gray,linewidth=2.2pt](0.37,-0.52) 
\psdot[linecolor=red,linewidth=2.2pt](0.56,-0.32) 
\rput(0,0.06){\blue \bf BRAP4}
\rput(0.26,0){\blue \bf VALE3}
\rput(-0.56,0.06){SBSP3}
\rput(-1.20,0.06){CSMG3}
\rput(-1.91,0.06){SULA11}
\rput(-0.52,0.21){MMXM3}
\rput(-0.80,0.52){LLXL3}
\rput(-0.34,0.64){PINE4}
\rput(0,0.70){PSSA3}
\rput(0.34,0.64){SLCE3}
\rput(0.54,0.38){TEMP3}
\rput(-0.59,-0.33){ABCB4}
\rput(-0.39,-0.53){ALLL3}
\rput(-0.17,-0.71){AMIL3}
\rput(0.12,-0.72){GPCP3}
\rput(0.37,-0.58){HYPE3}
\rput(0.56,-0.38){OHLB3}
\end{pspicture}

\begin{center}
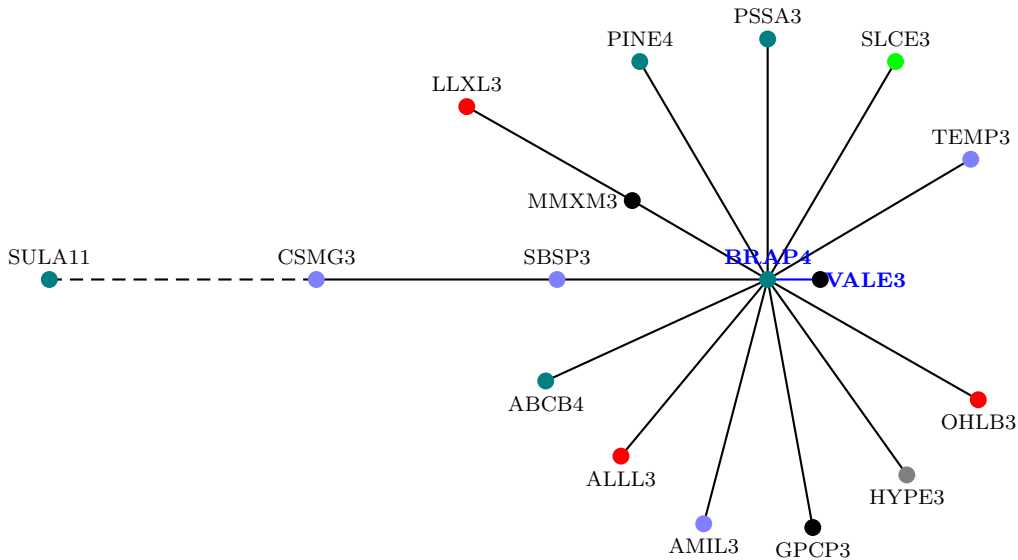
{\bf Figure 4.} Cluster around BRAP4.
\end{center}

As can be seen from the excerpt of the minimum spanning tree, it has strong ties with VALE3 (of Vale, a mining industry of which it has about 17\% of the stocks). It is immediately surrounded by 13 stocks, three of them of financial background (PINE4 and ABCB4 are both stocks of banks, PSSA3, of an insurance company), health (TEMP3 and AMIL3), sanitation (SBSP3, itself linked with stocks of another sanitation company, CSMG3), mining (VALE3 and MMXM3), petrochemistry (GPCP3), logistics (ALLL3, LLXL2, and OHLB3), and agribusiness (SLCE3). Indirectly and weakly connected (as shown by the dashed line) to this hub are the stocks of another insurance company (SULA11). This is a somewhat mixed cluster, with a variety of stocks orbiting the stocks of a very capitalized bank (Bradesco is the third major bank of Brazil).

The second and third clusters are actually the same, and they are represented in different figures in order to make it easier to discern the organization of the stocks arounds VALE3 and VALE5. Beginning with Figure 5, one can see a cluster formed around VALE3, which is strongly correlated with VALE5, as it would be expected. Vale is a mining company, and one can observe that connected to it, on its top, there is a collection of other stocks of mining companies (CSNA3, USIM3, and USIM5), of metalurgy (GGBR3, GGBR4, and GOAU4, all of them from the Gerdau group, and MAGG3), and of petrochemistry (UNIP6). Also connected, indirectly, to VALE3, are the stocks PETR3 and PETR4, of Petrobras, an oil, gas, and biofuel company which is one of the major in the world. Petrobras is responsible for a large amount of the volume traded in the Bovespa, and it is surprising to find it not as a hub itself in the minimum spanning tree representation.

A likely explanation is that the main product of Petrobras, petroil, is a comodity negotiated worldwide and so much more dependent on factors like the price of the oil barrel then on internal ones. This is enough to place Petrobras' stocks apart from the others.

At the lower part of Figure 5, one can see a cluster of stocks related with electricity distribution (ELET3 and ELET6, ELPL4, CPFE3, ENBR3, LIGT3, CPLE6, and CMIG3 and CMIG4), most sparsely correlated with one another. To the left, there is a cluster of stocks related with telecommunications (TNPL3 and TNPL4, TMAR5, and BRTO3 and BRTO4). There are also some more stocks, related with engineering and materials, logistics, food, heavy machinery, consumer goods, and health, scattered and not forming any particular cluster. Note that most of these stocks, arranged apparently at random, have weak connections, as shown by the dashed lines.

\begin{pspicture}(-8,-12.6)(1,9)
\psset{xunit=5,yunit=5} \scriptsize
\psline[linecolor=blue](0,0)(0.03,0) 
\psline[linecolor=blue](0,0)(-0.14,0) 
\psline(0,0)(0,0.34) 
\psline(0,0)(0.62,-0.16) 
\psline(0,0)(0.35,-0.35) 
\psline(0,0)(0,-0.58) 
\psline(0,0)(-0.41,-0.41) 
\psline[linestyle=dashed](0,0)(-0.68,-0.18) 
\psline(0,-0.58)(0,-0.68) 
\psline(0,-0.68)(0,-1.21) 
\psline(0,-1.21)(0,-1.79) 
\psline(0,-1.79)(0,-2.42) 
\psline(0,-1.79)(0.64,-1.79) 
\psline(0,-1.79)(-0.50,-1.79) 
\psline(-0.50,-1.79)(-0.78,-1.79) 
\psline(-0.78,-1.79)(-1.06,-1.79) 
\psline[linestyle=dashed](0.35,-0.35)(1.08,-0.35) 
\psline(0.35,-0.35)(0.50,-0.50) 
\psline(-0.41,-0.41)(-0.89,-0.89) 
\psline(-0.41,-0.41)(-1.04,-0.41) 
\psline(0,0.34)(-0.24,0.34) 
\psline(0,0.34)(0.60,0.34) 
\psline(0,0.34)(0,0.86) 
\psline(0,0.34)(0.18,0.52) 
\psline(0.60,0.34)(1.21,0.34) 
\psline(1.21,0.34)(1.53,0.89) 
\psline[linestyle=dashed](1.21,0.34)(1.57,-0.28) 
\psline(1.57,-0.28)(1.57,-0.11) 
\psline[linestyle=dashed](1.57,-0.28)(1.57,-1.05) 
\psline(1.57,-1.05)(1.57,-1.28) 
\psline[linestyle=dashed](1.53,0.89)(1.53,0.16) 
\psline[linestyle=dashed](1.53,0.89)(1.53,1.63) 
\psline(0.18,0.52)(0.26,0.60) 
\psline[linestyle=dashed](0.26,0.60)(0.78,1.12) 
\psline(0.26,0.60)(0.26,1.28) 
\psline(-0.24,0.34)(-0.78,0.20) 
\psline(-0.24,0.34)(-0.28,0.34) 
\psline(-0.24,0.34)(-0.29,0.39) 
\psline(-0.24,0.34)(-0.24,0.80) 
\psline(-0.24,0.80)(-0.24,0.85) 
\psline(-0.29,0.39)(-0.69,0.79) 
\psline[linestyle=dashed](-0.69,0.79)(-1.21,1.31) 
\psline[linestyle=dashed](-0.28,0.34)(-0.88,0.68) 
\psline(-0.28,0.34)(-0.86,0.34) 
\psline(-0.86,0.34)(-1.12,0.34) 
\psline(-0.86,0.34)(-1.26,0.23) 
\psline(-1.26,0.23)(-1.26,-0.19) 
\psline(-1.26,-0.19)(-1.26,-0.70) 
\psdot[linecolor=black,linewidth=2.2pt](0,0) 
\psdot[linecolor=black,linewidth=2.2pt](0,0.34) 
\psdot[linecolor=green,linewidth=2.2pt](0.62,-0.16) 
\psdot[linecolor=black,linewidth=2.2pt](0.03,0) 
\psdot[linecolor=magenta,linewidth=2.2pt](0.35,-0.35) 
\psdot[linecolor=orange,linewidth=2.2pt](0,-0.58) 
\psdot[linecolor=black,linewidth=2.2pt](-0.41,-0.41) 
\psdot[linecolor=blue,linewidth=2.2pt](-0.68,-0.18) 
\psdot[linecolor=darkgreen,linewidth=2.2pt](-0.14,0) 
\psdot[linecolor=orange,linewidth=2.2pt](0,-0.68) 
\psdot[linecolor=orange,linewidth=2.2pt](0,-1.21) 
\psdot[linecolor=orange,linewidth=2.2pt](0,-1.79) 
\psdot[linecolor=orange,linewidth=2.2pt](0,-2.42) 
\psdot[linecolor=orange,linewidth=2.2pt](0.64,-1.79) 
\psdot[linecolor=orange,linewidth=2.2pt](-0.50,-1.79) 
\psdot[linecolor=orange,linewidth=2.2pt](-0.78,-1.79) 
\psdot[linecolor=orange,linewidth=2.2pt](-1.06,-1.79) 
\psdot[linecolor=red,linewidth=2.2pt](1.08,-0.35) 
\psdot[linecolor=magenta,linewidth=2.2pt](0.50,-0.50) 
\psdot[linecolor=gray,linewidth=2.2pt](-0.89,-0.89) 
\psdot[linecolor=black,linewidth=2.2pt](-1.04,-0.41) 
\psdot[linecolor=black,linewidth=2.2pt](-0.24,0.34) 
\psdot[linecolor=black,linewidth=2.2pt](0.60,0.34) 
\psdot[linecolor=black,linewidth=2.2pt](0,0.86) 
\psdot[linecolor=black,linewidth=2.2pt](0.18,0.52) 
\psdot[linecolor=blue,linewidth=2.2pt](1.21,0.34) 
\psdot[linecolor=red,linewidth=2.2pt](1.53,0.89) 
\psdot[linecolor=black,linewidth=2.2pt](1.57,-0.28) 
\psdot[linecolor=black,linewidth=2.2pt](1.57,-0.11) 
\psdot[linecolor=blue,linewidth=2.2pt](1.57,-1.05) 
\psdot[linecolor=blue,linewidth=2.2pt](1.57,-1.28) 
\psdot[linecolor=gray,linewidth=2.2pt](1.53,0.16) 
\psdot[linecolor=red,linewidth=2.2pt](1.53,1.63) 
\psdot[linecolor=black,linewidth=2.2pt](0.26,0.60) 
\psdot[linecolor=lightyblue,linewidth=2.2pt](0.78,1.12) 
\psdot[linecolor=gray,linewidth=2.2pt](0.26,1.28) 
\psdot[linecolor=darkgreen,linewidth=2.2pt](-0.78,0.20) 
\psdot[linecolor=black,linewidth=2.2pt](-0.28,0.34) 
\psdot[linecolor=black,linewidth=2.2pt](-0.29,0.39) 
\psdot[linecolor=black,linewidth=2.2pt](-0.24,0.80) 
\psdot[linecolor=black,linewidth=2.2pt](-0.24,0.85) 
\psdot[linecolor=red,linewidth=2.2pt](-0.69,0.79) 
\psdot[linecolor=gray,linewidth=2.2pt](-1.21,1.31) 
\psdot[linecolor=blue,linewidth=2.2pt](-0.88,0.68) 
\psdot[linecolor=blue,linewidth=2.2pt](-0.86,0.34) 
\psdot[linecolor=blue,linewidth=2.2pt](-1.12,0.34) 
\psdot[linecolor=blue,linewidth=2.2pt](-1.26,0.23) 
\psdot[linecolor=blue,linewidth=2.2pt](-1.26,-0.19) 
\psdot[linecolor=blue,linewidth=2.2pt](-1.26,-0.70) 
\rput(-0.12,0.06){\blue \bf VALE3}
\rput(0.12,0.28){CSNA3}
\rput(0.74,-0.16){BEEF3}
\rput(0.16,0){\blue \bf VALE5}
\rput(0.23,-0.35){INEP4}
\rput(-0.12,-0.58){ELET3}
\rput(-0.29,-0.41){FESA4}
\rput(-0.80,-0.18){IGBR3}
\rput(-0.28,0){\blue \bf BRAP4}
\rput(-0.12,-0.68){ELET6}
\rput(-0.12,-1.21){ELPL4}
\rput(-0.12,-1.73){CPFE3}
\rput(0,-2.48){LIGT3}
\rput(0.64,-1.73){ENBR3}
\rput(-0.50,-1.73){CPLE6}
\rput(-0.78,-1.73){CMIG4}
\rput(-1.06,-1.73){CMIG3}
\rput(1.20,-0.35){PLAS3}
\rput(0.50,-0.56){INEP3}
\rput(-0.89,-0.95){KROT11}
\rput(-1.04,-0.35){CNFB4}
\rput(-0.12,0.40){GGBR4}
\rput(0.60,0.40){UNIP6}
\rput(0,0.92){MAGG3}
\rput(0.30,0.48){USIM5}
\rput(1.33,0.34){INET3}
\rput(1.65,0.89){IENG5}
\rput(1.71,-0.28){RPMG3}
\rput(1.71,-0.11){RPMG4}
\rput(1.69,-1.05){TELB3}
\rput(1.57,-1.34){TELB4}
\rput(1.65,0.16){TOYB3}
\rput(1.53,1.69){TGMA3}
\rput(0.38,0.60){USIM3}
\rput(0.78,1.18){CREM3}
\rput(0.26,1.34){PFRM3}
\rput(-0.90,0.20){BICB4}
\rput(-0.28,0.28){GGBR3}
\rput(-0.43,0.39){GOAU4}
\rput(-0.38,0.80){PETR4}
\rput(-0.24,0.91){PETR3}
\rput(-0.83,0.79){PMAM3}
\rput(-1.21,1.37){TEKA4}
\rput(-1.00,0.68){UOLL4}
\rput(-0.86,0.40){TNLP4}
\rput(-1.12,0.40){TNLP3}
\rput(-1.41,0.23){TMAR5}
\rput(-1.40,-0.19){BRTO4}
\rput(-1.26,-0.76){BRTO3}
\end{pspicture}

\begin{center}
{\bf Figure 5.} Cluster around VALE3.
\end{center}

\newpage 

Figure 6 shows the cluster around VALE5, which is the same cluster as the one of Figure 5, since VALE3 and VALE5 are intimately connected. At the right of VALE5, one can see a cluster of stocks belonging to companies related to agriculture (FFTL4), food (BRFS3, MRFG3, and RNAR3), paper (the sequence made by FIBR3, SUZB5, and KLBN4), and sugar and ethanol (SMTO3). There are also stocks belonging to other companies, but they apparently do not form clusters.

\begin{pspicture}(-8,-6)(1,9)
\psset{xunit=5,yunit=5} \scriptsize
\psline[linecolor=blue](0,0)(-0.03,0) 
\psline[linecolor=blue](0,0)(0.36,0) 
\psline(0,0)(0.30,0.30) 
\psline(0,0)(-0.42,-0.42) 
\psline(0,0)(0,0.56) 
\psline(0,0)(-0.31,0.54) 
\psline[linestyle=dashed](0,0)(-0.62,0.36) 
\psline(0,0)(0.54,-0.31) 
\psline[linestyle=dashed](0,0)(0.34,-0.60) 
\psline(0,0)(0,-0.64) 
\psline(-0.42,-0.42)(-1.08,-0.42) 
\psline(-0.42,-0.42)(-0.42,-1.09) 
\psline[linestyle=dashed](-0.42,-0.42)(-0.91,-0.91) 
\psline(-0.42,-0.42)(-0.82,-0.03) 
\psline(-0.82,-0.03)(-1.27,0.42) 
\psline(0.30,0.30)(0.30,0.87) 
\psline(0.30,0.30)(0.61,0.30) 
\psline[linestyle=dashed](0.30,0.87)(0.30,1.57) 
\psline(0.30,0.87)(-0.37,0.87) 
\psline(-0.37,0.87)(-1.04,0.87) 
\psline(0.61,0.30)(0.61,0.86) 
\psline(0.61,0.30)(1.03,0.72) 
\psline(0.61,0.30)(1.11,0.30) 
\psline(0.61,0.30)(1.04,-0.13) 
\psline[linestyle=dashed](0.61,0.86)(0.61,1.60) 
\psdot[linecolor=black,linewidth=2.2pt](0,0) 
\psdot[linecolor=black,linewidth=2.2pt](-0.03,0) 
\psdot[linecolor=darkgreen,linewidth=2.2pt](0.36,0) 
\psdot[linecolor=green,linewidth=2.2pt](0.30,0.30) 
\psdot[linecolor=red,linewidth=2.2pt](-0.42,-0.42) 
\psdot[linecolor=magenta,linewidth=2.2pt](0,0.56) 
\psdot[linecolor=blue,linewidth=2.2pt](-0.31,0.54) 
\psdot[linecolor=blue,linewidth=2.2pt](-0.62,0.36) 
\psdot[linecolor=black,linewidth=2.2pt](0.54,-0.31) 
\psdot[linecolor=gray,linewidth=2.2pt](0.34,-0.60) 
\psdot[linecolor=brown,linewidth=2.2pt](0,-0.64) 
\psdot[linecolor=gray,linewidth=2.2pt](-1.08,-0.42) 
\psdot[linecolor=green,linewidth=2.2pt](-0.42,-1.09) 
\psdot[linecolor=darkgreen,linewidth=2.2pt](-0.91,-0.91) 
\psdot[linecolor=brown,linewidth=2.2pt](-0.82,-0.03) 
\psdot[linecolor=brown,linewidth=2.2pt](-1.27,0.42) 
\psdot[linecolor=green,linewidth=2.2pt](0.30,0.87) 
\psdot[linecolor=green,linewidth=2.2pt](0.61,0.30) 
\psdot[linecolor=darkgreen,linewidth=2.2pt](0.30,1.57) 
\psdot[linecolor=blue,linewidth=2.2pt](-0.37,0.87) 
\psdot[linecolor=green,linewidth=2.2pt](-1.04,0.87) 
\psdot[linecolor=green,linewidth=2.2pt](0.61,0.86) 
\psdot[linecolor=magenta,linewidth=2.2pt](0.61,1.60) 
\psdot[linecolor=green,linewidth=2.2pt](1.03,0.72) 
\psdot[linecolor=green,linewidth=2.2pt](1.11,0.30) 
\psdot[linecolor=green,linewidth=2.2pt](1.04,-0.13) 
\rput(0,0.06){\blue \bf VALE5}
\rput(-0.18,0){\blue \bf VALE3}
\rput(0.51,0){\blue \bf BBDC4}
\rput(0.18,0.30){FIBR3}
\rput(-0.28,-0.42){LOGN3}
\rput(0,0.62){KEPL3}
\rput(-0.31,0.60){NETC4}
\rput(-0.62,0.42){BEMA3}
\rput(0.54,-0.37){ECOD3}
\rput(0.34,-0.66){ESTR4}
\rput(0,-0.70){ETER3}
\rput(-1.22,-0.42){SGPS3}
\rput(-0.42,-1.15){MNPR3}
\rput(-0.91,-0.97){VLID3}
\rput(-0.96,-0.02){TCSA3}
\rput(-1.27,0.48){HBOR3}
\rput(0.18,0.93){FFTL4}
\rput(0.49,0.36){SUZB5}
\rput(0.30,1.63){CARD3}
\rput(-0.37,0.93){CTAX4}
\rput(-1.04,0.93){RNAR3}
\rput(0.76,0.86){SMTO3}
\rput(0.74,1.60){PRVI3}
\rput(1.03,0.78){MRFG3}
\rput(1.11,0.36){KLBN4}
\rput(1.04,-0.19){BRFS3}
\end{pspicture}

\begin{center}
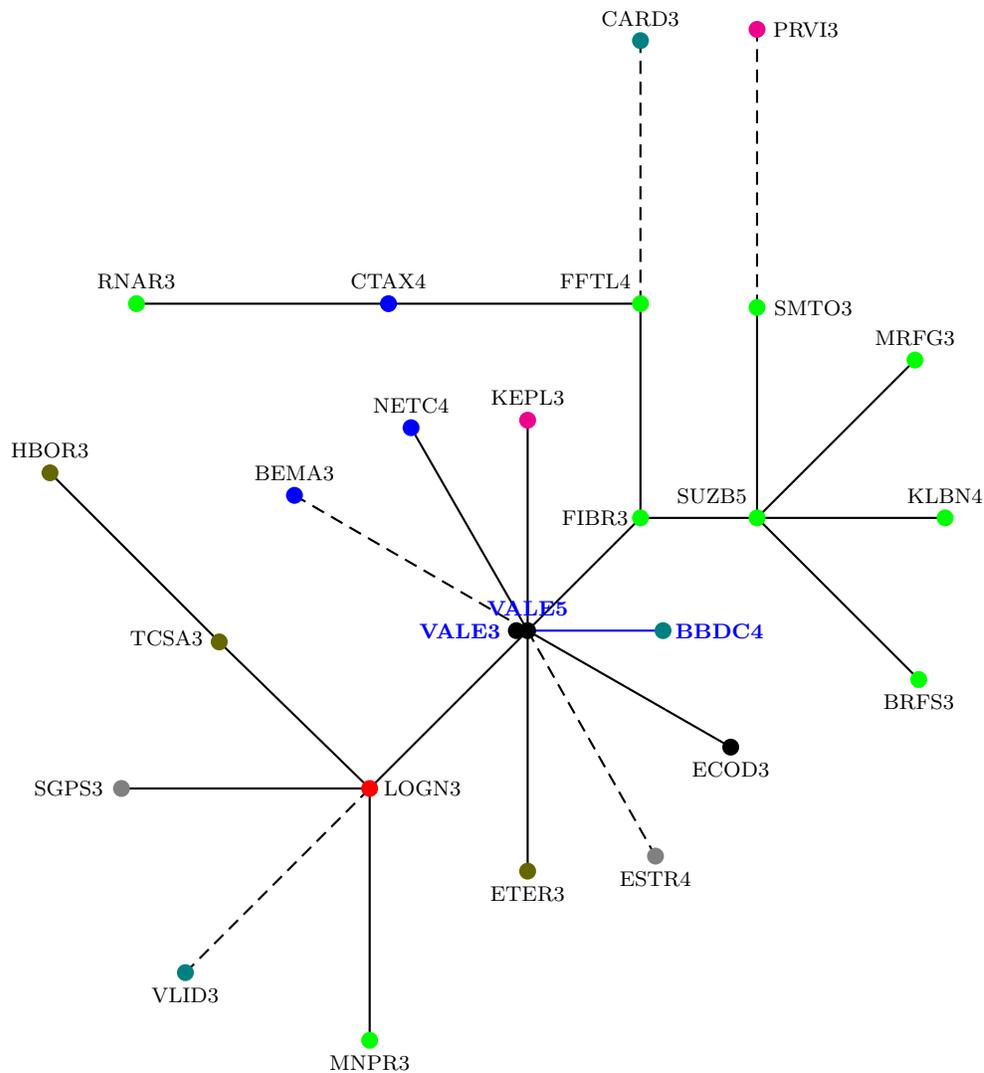
{\bf Figure 6.} Cluster around VALE5.
\end{center}

\newpage 

The cluster in Figure 7 is built around BBDC4 (closely connected wit BBDC3), which are stocks of Bank Bradesco. It is a denser cluster, comprised in its center by stocks related with banks (ITUB3, ITUB4, and ITSA4, all related with Bank Itau, the second largest in Brazil, SANB3, SANB4, and SANB11, stocks from Bank Santander, BBAS3, stocks of the Bank of Brazil, the largest in Brazil, and BRSR6) or with investment and finance companies (GPIV11, RDCD3), and BVMF3, wich are stocks of BM\&F-Bovespa itself. To the left, there are two stocks of COSAN (CSAN3 and CZLT11), a food, sugar, and ethanol company. To the top and right, there is another cluster, of cyclic consumer goods: beverages (AMBV3 and AMBV4), tobacco (CRUZ3), pharmaceuticals (DROG3), and sandals (ALPA4). Also to be noticed are the sequences RDCD3-CIEL3, both related with companies that operate and sell credit and debit card terminals, and GOLL4-TAMM4, both related with air transport companies. Another small cluster comprises the stocks TCSL3 and TCSL4, and VIVO4, related with mobile telephony companies. Other stocks scatter around the main network without forming any discernible subnetwork.

\begin{pspicture}(-8,-6.9)(1,8)
\psset{xunit=5,yunit=5} \scriptsize
\psline[linecolor=blue](0,0)(-0.36,0) 
\psline[linecolor=blue](0,0)(0.43,0) 
\psline(0,0)(0,-0.12) 
\psline(0,0)(0.30,0.05) 
\psline(0,-0.12)(0,-0.20) 
\psline(0,-0.20)(0,-0.78) 
\psline(0,-0.78)(-0.32,-1.32) 
\psline(0,-0.78)(0.32,-1.34) 
\psline(0,-0.20)(0.26,-0.65) 
\psline(0,-0.20)(0.48,-0.48) 
\psline(0,-0.20)(-0.28,-0.69) 
\psline[linestyle=dashed](0.26,-0.65)(0.60,-1.25) 
\psline(-0.28,-0.69)(-0.47,-1.02) 
\psline(0,-0.12)(-0.16,-0.15) 
\psline(0,0)(0.46,-0.08) 
\psline(0.46,-0.08)(0.79,-0.14) 
\psline(0,0)(0.64,-0.23) 
\psline(0,-0.12)(0.66,-0.36) 
\psline(0,-0.12)(-0.42,-0.54) 
\psline(0,0)(-0.59,-0.10) 
\psline(-0.59,-0.10)(-1.15,-0.20) 
\psline(0,0)(-0.46,0.08) 
\psline(0,0)(0.49,0.18) 
\psline(0,0)(0.42,0.37) 
\psline(0,0)(0,0.08) 
\psline(0,0)(-0.35,0.13) 
\psline(0.49,0.18)(0.85,0.18) 
\psline(0.49,0.18)(0.98,0.00) 
\psline[linestyle=dashed](0.42,0.37)(1.03,0.72) 
\psline(0.42,0.37)(0.67,0.37) 
\psline[linestyle=dashed](0.67,0.37)(1.37,0.37) 
\psline[linestyle=dashed](0.42,0.37)(0.78,0.98) 
\psline(-0.35,0.13)(-0.93,0.13) 
\psline(-0.93,0.13)(-1.29,0.13) 
\psline(-0.35,0.13)(-0.98,0.36) 
\psline(-0.35,0.13)(-0.75,0.46) 
\psline(0,0.08)(0,0.67) 
\psline[linestyle=dashed](0,0.67)(0,1.36) 
\psline(0,0.08)(0.32,0.62) 
\psline[linestyle=dashed](0.32,0.62)(0.32,1.36) 
\psline(0,0.08)(-0.32,0.63) 
\psline(-0.32,0.63)(-0.32,1.30) 
\psdot[linecolor=darkgreen,linewidth=2.2pt](0,0) 
\psdot[linecolor=black,linewidth=2.2pt](-0.36,0) 
\psdot[linecolor=brown,linewidth=2.2pt](0.43,0) 
\psdot[linecolor=darkgreen,linewidth=2.2pt](0,-0.12) 
\psdot[linecolor=darkgreen,linewidth=2.2pt](0.30,0.05) 
\psdot[linecolor=darkgreen,linewidth=2.2pt](0,-0.20) 
\psdot[linecolor=green,linewidth=2.2pt](0,-0.78) 
\psdot[linecolor=magenta,linewidth=2.2pt](-0.32,-1.32) 
\psdot[linecolor=blue,linewidth=2.2pt](0.32,-1.34) 
\psdot[linecolor=black,linewidth=2.2pt](0.26,-0.65) 
\psdot[linecolor=magenta,linewidth=2.2pt](0.48,-0.48) 
\psdot[linecolor=darkgreen,linewidth=2.2pt](-0.28,-0.69) 
\psdot[linecolor=brown,linewidth=2.2pt](0.60,-1.25) 
\psdot[linecolor=darkgreen,linewidth=2.2pt](-0.47,-1.02) 
\psdot[linecolor=darkgreen,linewidth=2.2pt](-0.16,-0.15) 
\psdot[linecolor=red,linewidth=2.2pt](0.46,-0.08) 
\psdot[linecolor=red,linewidth=2.2pt](0.79,-0.14) 
\psdot[linecolor=brown,linewidth=2.2pt](0.64,-0.23) 
\psdot[linecolor=darkgreen,linewidth=2.2pt](0.66,-0.36) 
\psdot[linecolor=red,linewidth=2.2pt](-0.42,-0.54) 
\psdot[linecolor=blue,linewidth=2.2pt](-0.59,-0.10) 
\psdot[linecolor=brown,linewidth=2.2pt](-1.15,-0.20) 
\psdot[linecolor=darkgreen,linewidth=2.2pt](-0.46,0.08) 
\psdot[linecolor=blue,linewidth=2.2pt](0.49,0.18) 
\psdot[linecolor=gray,linewidth=2.2pt](0.42,0.37) 
\psdot[linecolor=darkgreen,linewidth=2.2pt](0,0.08) 
\psdot[linecolor=darkgreen,linewidth=2.2pt](-0.35,0.13) 
\psdot[linecolor=blue,linewidth=2.2pt](0.85,0.18) 
\psdot[linecolor=blue,linewidth=2.2pt](0.98,0.00) 
\psdot[linecolor=gray,linewidth=2.2pt](1.03,0.72) 
\psdot[linecolor=gray,linewidth=2.2pt](0.67,0.37) 
\psdot[linecolor=gray,linewidth=2.2pt](1.37,0.37) 
\psdot[linecolor=gray,linewidth=2.2pt](0.78,0.98) 
\psdot[linecolor=green,linewidth=2.2pt](-0.93,0.13) 
\psdot[linecolor=green,linewidth=2.2pt](-1.29,0.13) 
\psdot[linecolor=darkgreen,linewidth=2.2pt](-0.98,0.36) 
\psdot[linecolor=darkgreen,linewidth=2.2pt](-0.75,0.46) 
\psdot[linecolor=darkgreen,linewidth=2.2pt](0,0.67) 
\psdot[linecolor=orange,linewidth=2.2pt](0,1.36) 
\psdot[linecolor=green,linewidth=2.2pt](0.32,0.62) 
\psdot[linecolor=orange,linewidth=2.2pt](0.32,1.36) 
\psdot[linecolor=blue,linewidth=2.2pt](-0.32,0.63) 
\psdot[linecolor=blue,linewidth=2.2pt](-0.32,1.30) 
\rput(0,-0.06){\blue \bf BBDC4}
\rput(-0.51,0){\blue \bf VALE5}
\rput(0.58,0){\blue \bf GFSA3}
\rput(0.12,-0.12){ITUB4}
\rput(0.42,0.05){BBAS3}
\rput(0.12,-0.20){ITSA4}
\rput(0.12,-0.78){JBSS3}
\rput(-0.32,-1.38){HAGA4}
\rput(0.32,-1.40){MLFT4}
\rput(0.40,-0.65){OGXP3}
\rput(0.48,-0.54){EMBR3}
\rput(-0.42,-0.69){RDCD3}
\rput(0.60,-1.31){TCNO3}
\rput(-0.47,-1.08){CIEL3}
\rput(-0.28,-0.15){ITUB3}
\rput(0.46,-0.14){GOLL4}
\rput(0.79,-0.20){TAMM4}
\rput(0.64,-0.29){GSHP3}
\rput(0.66,-0.42){GPIV11}
\rput(-0.54,-0.54){FRAS4}
\rput(-0.59,-0.16){POSI3}
\rput(-1.15,-0.26){BBRK3}
\rput(-0.60,0.08){BVMF3}
\rput(0.49,0.24){TCSL4}
\rput(0.27,0.37){AMBV4}
\rput(0,0.14){BBDC3}
\rput(-0.25,0.19){SANB11}
\rput(0.97,0.18){TCSL3}
\rput(0.98,-0.06){VIVO4}
\rput(1.03,0.78){CRUZ3}
\rput(0.67,0.43){AMBV3}
\rput(1.37,0.43){DROG3}
\rput(0.78,1.04){ALPA4}
\rput(-0.93,0.19){CZLT11}
\rput(-1.29,0.19){CSAN3}
\rput(-1.12,0.36){SANB4}
\rput(-0.89,0.46){SANB3}
\rput(-0.14,0.67){BRSR6}
\rput(0,1.42){COCE5}
\rput(0.21,0.62){EUCA4}
\rput(0.32,1.42){CLSC6}
\rput(-0.46,0.63){TLPP4}
\rput(-0.32,1.36){TLPP3}
\end{pspicture}

\begin{center}
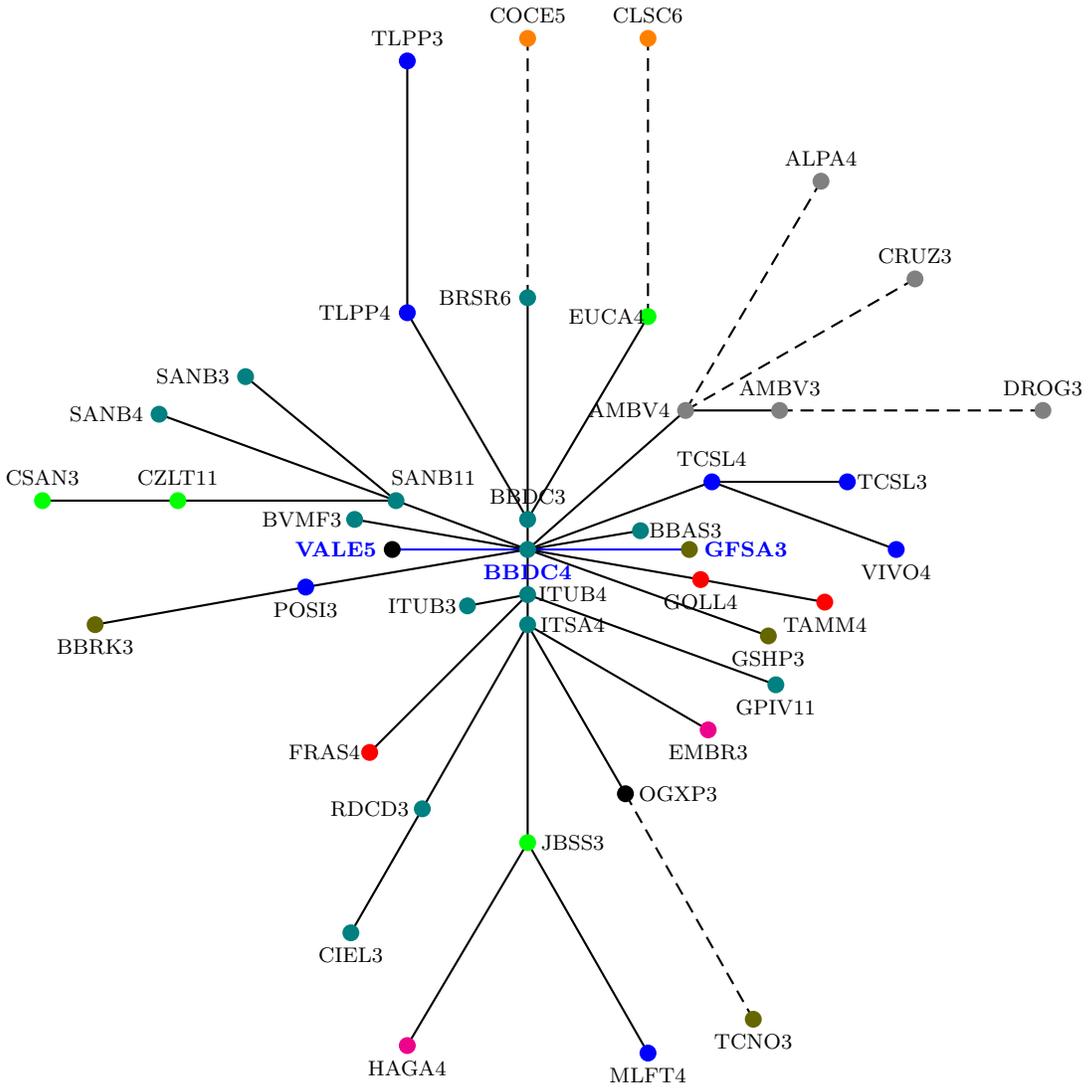
{Figure 7.} Cluster around BBDC4.
\end{center}

\newpage 

The next cluster, represented in Figure 8, is centered around a construction and real state company, Gafisa (GFSA3), and is composed mostly by the stocks of other construction and real state companies (PDGR3, CYRE3, CRDE3, EVEN3, INPR3, JFEN3, and LPSB3), and by stocks of consumer goods companies (LAME3 and LAME4, LREN3, BTOW3, NATU3, AMAR3, HGTX3, and MTIG4). There is also a small cluster (lower part the figure) of stocks of electricity distribution companies (GETI3 and GETI4, and TBLE3).

\begin{pspicture}(-8,-8.7)(1,7)
\psset{xunit=5,yunit=5} \scriptsize
\psline[linecolor=blue](0,0)(-0.43,0) 
\psline[linecolor=blue](0,0)(0.30,0) 
\psline(0,0)(-0.40,0.23) 
\psline(0,0)(0.20,0.12) 
\psline(0,0)(0,0.53) 
\psline(0,0)(0,-0.42) 
\psline(-0.40,0.23)(-0.90,0.52) 
\psline(-0.90,0.52)(-1.50,0.86) 
\psline(0.20,0.12)(0.78,0.12) 
\psline(0.20,0.12)(0.70,0.41) 
\psline(0,0.53)(0.50,0.82) 
\psline(0.50,0.82)(1.00,1.11) 
\psline(0,0.53)(0.25,0.96) 
\psline(0,0.53)(-0.21,0.89) 
\psline(0,0.53)(-0.38,0.75) 
\psline(-0.38,0.75)(-0.91,1.06) 
\psline(0,-0.42)(-0.59,-0.42) 
\psline(0,-0.42)(0.61,-0.42) 
\psline(0,-0.42)(-0.55,-0.74) 
\psline(0,-0.42)(0.48,-0.70) 
\psline(0.48,-0.70)(1.06,-1.04) 
\psline(0.48,-0.70)(-0.10,-1.04) 
\psline(-0.10,-1.04)(0.27,-1.26) 
\psline[linestyle=dashed](-0.10,-1.04)(-0.72,-1.40) 
\psline(0.61,-0.42)(1.26,-0.42) 
\psline[linestyle=dashed](0.61,-0.42)(1.22,-0.78) 
\psline(0.61,-0.42)(1.18,-0.09) 
\psline(1.18,-0.09)(1.18,0.58) 
\psline(-0.59,-0.42)(-1.25,-0.42) 
\psline[linestyle=dashed](-1.25,-0.42)(-1.25,0.28) 
\psline(-1.25,-0.42)(-1.25,-1.06) 
\psline(-1.25,-1.06)(-1.25,-1.72) 
\psdot[linecolor=brown,linewidth=2.2pt](0,0) 
\psdot[linecolor=darkgreen,linewidth=2.2pt](-0.43,0) 
\psdot[linecolor=brown,linewidth=2.2pt](0.30,0) 
\psdot[linecolor=brown,linewidth=2.2pt](-0.40,0.23) 
\psdot[linecolor=brown,linewidth=2.2pt](0.20,0.12) 
\psdot[linecolor=gray,linewidth=2.2pt](0,0.53) 
\psdot[linecolor=brown,linewidth=2.2pt](0,-0.42) 
\psdot[linecolor=green,linewidth=2.2pt](-0.90,0.52) 
\psdot[linecolor=darkgreen,linewidth=2.2pt](-1.50,0.86) 
\psdot[linecolor=brown,linewidth=2.2pt](0.78,0.12) 
\psdot[linecolor=red,linewidth=2.2pt](0.70,0.41) 
\psdot[linecolor=blue,linewidth=2.2pt](0.50,0.82) 
\psdot[linecolor=magenta,linewidth=2.2pt](1.00,1.11) 
\psdot[linecolor=gray,linewidth=2.2pt](0.25,0.96) 
\psdot[linecolor=gray,linewidth=2.2pt](-0.21,0.89) 
\psdot[linecolor=gray,linewidth=2.2pt](-0.38,0.75) 
\psdot[linecolor=gray,linewidth=2.2pt](-0.91,1.06) 
\psdot[linecolor=darkgreen,linewidth=2.2pt](-0.59,-0.42) 
\psdot[linecolor=red,linewidth=2.2pt](0.61,-0.42) 
\psdot[linecolor=magenta,linewidth=2.2pt](-0.55,-0.74) 
\psdot[linecolor=brown,linewidth=2.2pt](0.48,-0.70) 
\psdot[linecolor=gray,linewidth=2.2pt](1.06,-1.04) 
\psdot[linecolor=orange,linewidth=2.2pt](-0.10,-1.04) 
\psdot[linecolor=orange,linewidth=2.2pt](0.27,-1.26) 
\psdot[linecolor=orange,linewidth=2.2pt](-0.72,-1.40) 
\psdot[linecolor=darkgreen,linewidth=2.2pt](1.26,-0.42) 
\psdot[linecolor=brown,linewidth=2.2pt](1.22,-0.78) 
\psdot[linecolor=gray,linewidth=2.2pt](1.18,-0.09) 
\psdot[linecolor=gray,linewidth=2.2pt](1.18,0.58) 
\psdot[linecolor=brown,linewidth=2.2pt](-1.25,-0.42) 
\psdot[linecolor=red,linewidth=2.2pt](-1.25,0.28) 
\psdot[linecolor=gray,linewidth=2.2pt](-1.25,-1.06) 
\psdot[linecolor=lightyblue,linewidth=2.2pt](-1.25,-1.72) 
\rput(-0.14,-0.06){\blue \bf GFSA3}
\rput(-0.59,0){\blue \bf BBDC4}
\rput(0.46,0){\blue \bf PDGR3}
\rput(-0.28,0.29){EVEN3}
\rput(0.12,0.18){CYRE3}
\rput(-0.14,0.51){LAME4}
\rput(-0.14,-0.36){BISA3}
\rput(-0.80,0.58){FHER3}
\rput(-1.50,0.92){BTTL4}
\rput(0.78,0.18){CRDE3}
\rput(0.70,0.47){CCRO3}
\rput(0.55,0.76){IDNT3}
\rput(1.00,1.17){LUPA3}
\rput(0.25,1.02){BTOW3}
\rput(-0.21,0.95){LAME3}
\rput(-0.44,0.69){LREN3}
\rput(-0.91,1.12){NATU3}
\rput(-0.59,-0.36){BPNM4}
\rput(0.51,-0.36){POMO4}
\rput(-0.55,-0.80){FJTA4}
\rput(0.52,-0.64){INPR3}
\rput(1.06,-1.10){MTIG4}
\rput(-0.13,-0.98){GETI4}
\rput(0.27,-1.32){GETI3}
\rput(-0.72,-1.46){TBLE3}
\rput(1.26,-0.48){SFSA4}
\rput(1.22,-0.84){JFEN3}
\rput(1.32,-0.09){AMAR3}
\rput(1.18,0.64){HGTX3}
\rput(-1.39,-0.42){LPSB3}
\rput(-1.25,0.34){TPIS3}
\rput(-1.38,-1.06){AEDU3}
\rput(-1.25,-1.78){DASA3}
\end{pspicture}

\begin{center}
{\bf Figure8.} Cluster around GFSA3.
\end{center}

\newpage 

The last cluster, Figure 9, is also centered around the stock of a construction and real state company, PDG Realty (PDGR3), and is composed mostly of stocks of other construction and real state companies (GFSA3, EZTC3, RSID3, MRVE3, JHSF3, and CCIM3), real state management of shopping centers (BRML3, MULT3, and IGTA3), and of building materials (DTEX3). Other stocks belong to electricity, logistics and transportation, consumer goods and stocks of other types of companies.

\begin{pspicture}(-7,-7.7)(1,7.3)
\psset{xunit=5,yunit=5} \scriptsize
\psline[linecolor=blue](0,0)(-0.30,0) 
\psline(0,0)(0.47,0) 
\psline(0.47,0)(1.14,0) 
\psline(0,0)(0.48,0.28) 
\psline(0,0)(0.52,-0.30) 
\psline(0,0)(0,-0.22) 
\psline(0,0)(0,0.26) 
\psline(0,0.26)(0.52,0.56) 
\psline(0,0.26)(-0.50,0.55) 
\psline(0.52,0.56)(1.14,0.56) 
\psline[linestyle=dashed](1.14,0.56)(1.14,1.33) 
\psline(-0.50,0.55)(-1.05,0.55) 
\psline(-1.05,0.55)(-1.05,-0.10) 
\psline[linestyle=dashed](-1.05,-0.10)(-1.05,-0.85) 
\psline(-0.50,0.55)(-0.98,1.03) 
\psline(-0.50,0.55)(-0.50,1.27) 
\psline(-0.50,0.55)(-0.03,1.02) 
\psline(-0.50,0.55)(-0.16,0.55) 
\psline(0,-0.22)(0,-0.83) 
\psline(0,-0.22)(-0.30,-0.73) 
\psline(0,-0.22)(0.28,-0.70) 
\psline[linestyle=dashed](-0.30,-0.73)(-0.66,-1.35) 
\psline(0,-0.83)(0,-1.44) 
\psline(0.28,-0.70)(0.57,-1.20) 
\psline(0.28,-0.70)(0.92,-0.70) 
\psline[linestyle=dashed](0.28,-0.70)(0.89,-0.35) 
\psline(0.92,-0.70)(1.53,-0.70) 
\psline(0.92,-0.70)(1.49,-0.37) 
\psline[linestyle=dashed](1.49,-0.37)(1.49,0.40) 
\psline(0.57,-1.20)(1.07,-1.20) 
\psline(1.07,-1.20)(1.65,-1.20) 
\psdot[linecolor=brown,linewidth=2.2pt](0,0) 
\psdot[linecolor=brown,linewidth=2.2pt](-0.30,0) 
\psdot[linecolor=brown,linewidth=2.2pt](0.47,0) 
\psdot[linecolor=orange,linewidth=2.2pt](1.14,0) 
\psdot[linecolor=red,linewidth=2.2pt](0.48,0.28) 
\psdot[linecolor=orange,linewidth=2.2pt](0.52,-0.30) 
\psdot[linecolor=brown,linewidth=2.2pt](0,-0.22) 
\psdot[linecolor=brown,linewidth=2.2pt](0,0.26) 
\psdot[linecolor=orange,linewidth=2.2pt](0.52,0.56) 
\psdot[linecolor=brown,linewidth=2.2pt](-0.50,0.55) 
\psdot[linecolor=orange,linewidth=2.2pt](1.14,0.56) 
\psdot[linecolor=gray,linewidth=2.2pt](1.14,1.33) 
\psdot[linecolor=brown,linewidth=2.2pt](-1.05,0.55) 
\psdot[linecolor=magenta,linewidth=2.2pt](-1.05,-0.10) 
\psdot[linecolor=gray,linewidth=2.2pt](-1.05,-0.85) 
\psdot[linecolor=gray,linewidth=2.2pt](-0.98,1.03) 
\psdot[linecolor=gray,linewidth=2.2pt](-0.50,1.27) 
\psdot[linecolor=lightyblue,linewidth=2.2pt](-0.03,1.02) 
\psdot[linecolor=darkgreen,linewidth=2.2pt](-0.16,0.55) 
\psdot[linecolor=black,linewidth=2.2pt](0,-0.83) 
\psdot[linecolor=gray,linewidth=2.2pt](-0.30,-0.73) 
\psdot[linecolor=brown,linewidth=2.2pt](0.28,-0.70) 
\psdot[linecolor=blue,linewidth=2.2pt](-0.66,-1.35) 
\psdot[linecolor=darkgreen,linewidth=2.2pt](0,-1.44) 
\psdot[linecolor=brown,linewidth=2.2pt](0.57,-1.20) 
\psdot[linecolor=red,linewidth=2.2pt](0.92,-0.70) 
\psdot[linecolor=lightyblue,linewidth=2.2pt](0.89,-0.35) 
\psdot[linecolor=red,linewidth=2.2pt](1.53,-0.70) 
\psdot[linecolor=green,linewidth=2.2pt](1.49,-0.37) 
\psdot[linecolor=darkgreen,linewidth=2.2pt](1.49,0.40) 
\psdot[linecolor=brown,linewidth=2.2pt](1.07,-1.20) 
\psdot[linecolor=brown,linewidth=2.2pt](1.65,-1.20) 
\rput(-0.15,0.06){\blue \bf PDGR3}
\rput(-0.45,0){\blue \bf GFSA3}
\rput(0.47,0.06){EZTC3}
\rput(1.27,0){EQTL3}
\rput(0.48,0.34){RENT3}
\rput(0.52,-0.36){MPXE3}
\rput(-0.15,-0.22){MRVE3}
\rput(-0.15,0.26){RSID3}
\rput(0.52,0.62){CESP6}
\rput(-0.53,0.49){JHSF3}
\rput(1.29,0.56){TRPL4}
\rput(1.14,1.39){MNDL4}
\rput(-1.18,0.55){CCIM3}
\rput(-1.19,-0.10){WEGE3}
\rput(-1.05,-0.91){TOYB4}
\rput(-0.98,1.09){SLED4}
\rput(-0.50,1.33){GRND3}
\rput(-0.03,1.08){FLRY3}
\rput(-0.16,0.61){CTIP3}
\rput(-0.15,-0.83){BRKM5}
\rput(-0.45,-0.73){PCAR5}
\rput(0.15,-0.70){DTEX3}
\rput(-0.66,-1.41){TOTS3}
\rput(0,-1.50){UGPA4}
\rput(0.57,-1.26){BRML3}
\rput(0.80,-0.64){MYPK3}
\rput(0.89,-0.29){ODPV3}
\rput(1.67,-0.70){RAPT4}
\rput(1.63,-0.37){MILK11}
\rput(1.49,0.46){JBDU4}
\rput(1.07,-1.14){MULT3}
\rput(1.65,-1.14){IGTA3}
\end{pspicture}

\begin{center}
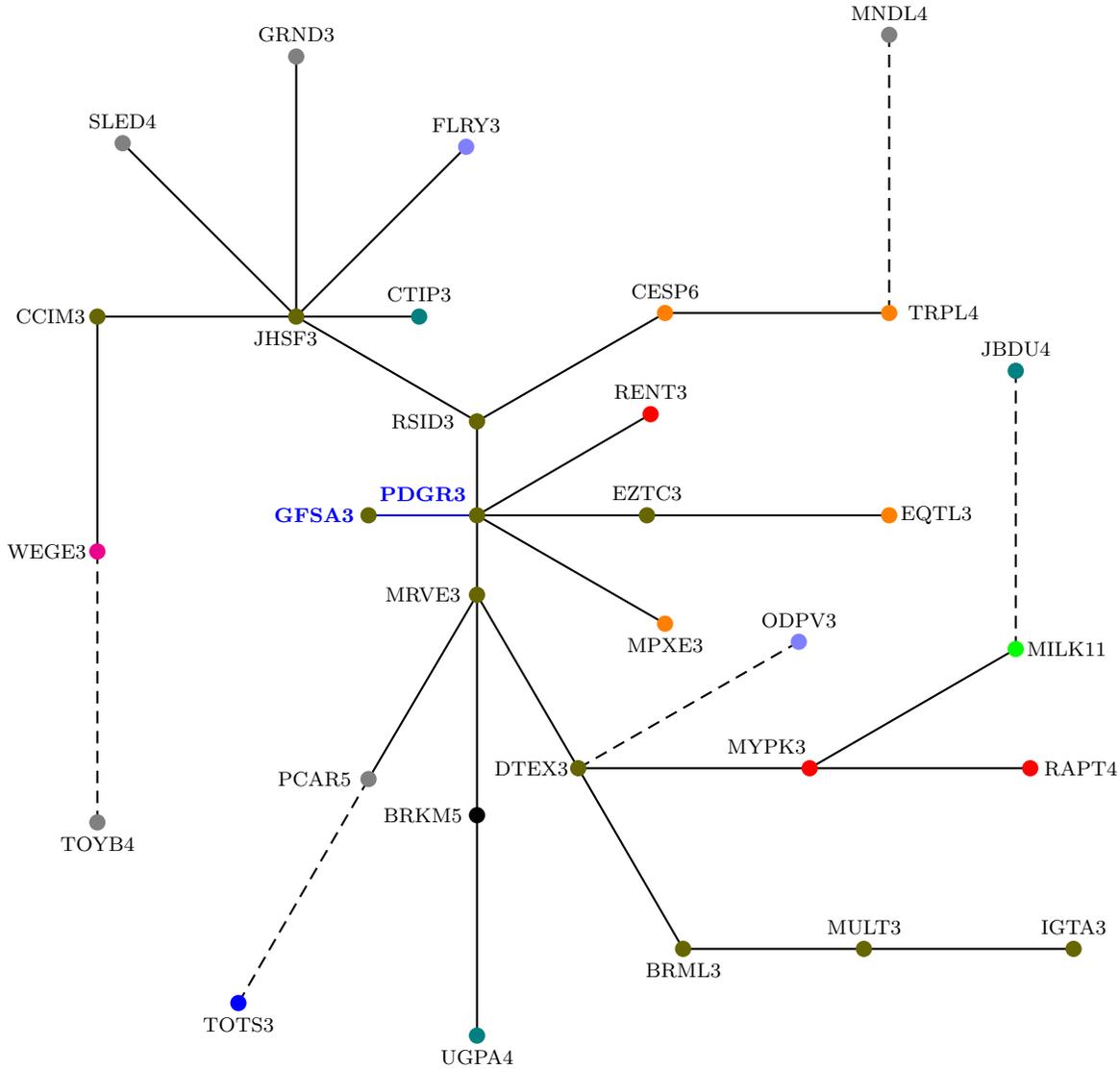
{\bf Figure 9.} Cluster around PDGR3.
\end{center}

\section{Centrality measures}

In network theory, the centrality of a node, or how influential the vertex is in the network, is an important measurement which is handled in a number of diferent ways. In what follows, I perform an analysis of the centrality of vertices in the network depicted by the MST of the last section according to five different definitions of centrality. I also do some analysis of the frequency distribution of each centrality measure in the network and which are the stocks that are more central according to each definition.

\vskip 0.3 cm

\noindent \it Node degree \rm

\vskip 0.2 cm

Next, I analize the node degree of the stocks in the network. The node degree of a node (stock in our case) is the number of connections it has in the network. Most of the stocks have low node degree, and some of them have a large number associated with them. The latter are called hubs, and are generally nodes that are more important in events that can change the network. Table 1 and Figure 10 show the node degree frequency distribution of the network of the BM\&F-Bovespa 2010.

\vskip 0.4 cm

\begin{minipage}{6 cm }
\[ \begin{array}{|c|c|} \hline \text{Node degree} & \text{Frequency} \\ \hline 1 & 105 \\ 2 & 45 \\ 3 & 18 \\ 4 & 7 \\ 5 & 8 \\ 6 & 3 \\ 7 & 0 \\ 8 & 1 \\ 9 & 0 \\ 10 & 1 \\ 11 & 0 \\ 12 & 1 \\ 13 & 1 \\ \hline \end{array} \]
{\bf Table 1.} Node degree distribution.
\end{minipage}
\begin{pspicture}(-0.5,3)(3.5,3)
\psset{xunit=0.5,yunit=0.05}
\pspolygon*[linecolor=lightred](0.5,0)(0.5,105)(1.5,105)(1.5,0)
\pspolygon*[linecolor=lightred](1.5,0)(1.5,45)(2.5,45)(2.5,0)
\pspolygon*[linecolor=lightred](2.5,0)(2.5,18)(3.5,18)(3.5,0)
\pspolygon*[linecolor=lightred](3.5,0)(3.5,7)(4.5,7)(4.5,0)
\pspolygon*[linecolor=lightred](4.5,0)(4.5,8)(5.5,8)(5.5,0)
\pspolygon*[linecolor=lightred](5.5,0)(5.5,3)(6.5,3)(6.5,0)
\pspolygon*[linecolor=lightred](6.5,0)(6.5,0)(7.5,0)(7.5,0)
\pspolygon*[linecolor=lightred](7.5,0)(7.5,1)(8.5,1)(8.5,0)
\pspolygon*[linecolor=lightred](8.5,0)(8.5,0)(9.5,0)(9.5,0)
\pspolygon*[linecolor=lightred](9.5,0)(9.5,1)(10.5,1)(10.5,0)
\pspolygon*[linecolor=lightred](10.5,0)(10.5,0)(11.5,0)(11.5,0)
\pspolygon*[linecolor=lightred](11.5,0)(11.5,1)(12.5,1)(12.5,0)
\pspolygon*[linecolor=lightred](12.5,0)(12.5,1)(13.5,1)(13.5,0)
\pspolygon[linecolor=red](0.5,0)(0.5,105)(1.5,105)(1.5,0)
\pspolygon[linecolor=red](1.5,0)(1.5,45)(2.5,45)(2.5,0)
\pspolygon[linecolor=red](2.5,0)(2.5,18)(3.5,18)(3.5,0)
\pspolygon[linecolor=red](3.5,0)(3.5,7)(4.5,7)(4.5,0)
\pspolygon[linecolor=red](4.5,0)(4.5,8)(5.5,8)(5.5,0)
\pspolygon[linecolor=red](5.5,0)(5.5,3)(6.5,3)(6.5,0)
\pspolygon[linecolor=red](6.5,0)(6.5,0)(7.5,0)(7.5,0)
\pspolygon[linecolor=red](7.5,0)(7.5,1)(8.5,1)(8.5,0)
\pspolygon[linecolor=red](8.5,0)(8.5,0)(9.5,0)(9.5,0)
\pspolygon[linecolor=red](9.5,0)(9.5,1)(10.5,1)(10.5,0)
\pspolygon[linecolor=red](10.5,0)(10.5,0)(11.5,0)(11.5,0)
\pspolygon[linecolor=red](11.5,0)(11.5,1)(12.5,1)(12.5,0)
\pspolygon[linecolor=red](12.5,0)(12.5,1)(13.5,1)(13.5,0)
%
%
\psline{->}(0,0)(15,0) \psline{->}(0,0)(0,120) \rput(17.5,0){node degree} \rput(2,120){frequency} \scriptsize \psline(1,-2)(1,2) \rput(1,-6){1} \psline(2,-2)(2,2) \rput(2,-6){2} \psline(3,-2)(3,2) \rput(3,-6){3} \psline(4,-2)(4,2) \rput(4,-6){4} \psline(5,-2)(5,2) \rput(5,-6){5} \psline(6,-2)(6,2) \rput(6,-6){6}  \psline(7,-2)(7,2) \rput(7,-6){7} \psline(8,-2)(8,2) \rput(8,-6){8} \psline(9,-2)(9,2) \rput(9,-6){9} \psline(10,-2)(10,2) \rput(10,-6){10} \psline(11,-2)(11,2) \rput(11,-6){11} \psline(12,-2)(12,2) \rput(12,-6){12} \psline(13,-2)(13,2) \rput(13,-6){13} \psline(-0.2,20)(0.2,20) \rput(-0.7,20){20} \psline(-0.2,40)(0.2,40) \rput(-0.7,40){40} \psline(-0.2,60)(0.2,60) \rput(-0.7,60){60} \psline(-0.2,80)(0.2,80) \rput(-0.7,80){80} \psline(-0.2,100)(0.2,100) \rput(-1,100){100}
\normalsize \rput(7.5,-14){{\bf Figure 10.} Node degree frequency distribution.}
\end{pspicture}

\vskip 0.5 cm

The stocks with highest node degree are BRAP4 (node degree 13), BBDC4 (node degree 12), VALE5 (node degree 10), and VALE3 (node degree 8).

\vskip 0.3 cm

\noindent \it Node strength \rm

\vskip 0.2 cm

The strength of a node is the sum of the correlations of the node with all other nodes to which it is connected. If $C$ is the matrix that stores the correlations between nodes that are linked in the minimum spanning tree, then the node strength is given by
\begin{equation}N_s^k=\sum_{i=1}^nC_{ik}+\sum_{j=1}^nC_{kj}\ ,\end{equation}where $C_{ij}$ is an element of matrix $C$.

In our network, the vertices with highest node strength are BBDC4 ($N_s=6.98$), BRAP4 ($N_s=5.37$), VALE5 ($N_s=4.72$), and VALE3 ($N_s=4.50$). The node strenght frequency distribution is shown in Figure 11.

\begin{pspicture}(-5,-0.5)(3.5,5.7)
\psset{xunit=0.8,yunit=0.05}
\pspolygon*[linecolor=lightred](0,0)(0,90)(0.5,90)(0.5,0)
\pspolygon*[linecolor=lightred](0.5,0)(0.5,47)(1,47)(1,0)
\pspolygon*[linecolor=lightred](1,0)(1,24)(1.5,24)(1.5,0)
\pspolygon*[linecolor=lightred](1.5,0)(1.5,13)(2,13)(2,0)
\pspolygon*[linecolor=lightred](2,0)(2,5)(2.5,5)(2.5,0)
\pspolygon*[linecolor=lightred](2.5,0)(2.5,2)(3,2)(3,0)
\pspolygon*[linecolor=lightred](3,0)(3,2)(3.5,2)(3.5,0)
\pspolygon*[linecolor=lightred](3.5,0)(3.5,3)(4,3)(4,0)
\pspolygon*[linecolor=lightred](4,0)(4,1)(4.5,1)(4.5,0)
\pspolygon*[linecolor=lightred](4.5,0)(4.5,1)(5,1)(5,0)
\pspolygon*[linecolor=lightred](5,0)(5,1)(5.5,1)(5.5,0)
\pspolygon*[linecolor=lightred](6.5,0)(6.5,1)(7,1)(7,0)
\pspolygon[linecolor=red](0,0)(0,90)(0.5,90)(0.5,0)
\pspolygon[linecolor=red](0.5,0)(0.5,47)(1,47)(1,0)
\pspolygon[linecolor=red](1,0)(1,24)(1.5,24)(1.5,0)
\pspolygon[linecolor=red](1.5,0)(1.5,13)(2,13)(2,0)
\pspolygon[linecolor=red](2,0)(2,5)(2.5,5)(2.5,0)
\pspolygon[linecolor=red](2.5,0)(2.5,2)(3,2)(3,0)
\pspolygon[linecolor=red](3,0)(3,2)(3.5,2)(3.5,0)
\pspolygon[linecolor=red](3.5,0)(3.5,3)(4,3)(4,0)
\pspolygon[linecolor=red](4,0)(4,1)(4.5,1)(4.5,0)
\pspolygon[linecolor=red](4.5,0)(4.5,1)(5,1)(5,0)
\pspolygon[linecolor=red](5,0)(5,1)(5.5,1)(5.5,0)
\pspolygon[linecolor=red](6.5,0)(6.5,1)(7,1)(7,0)
%
%
\psline{->}(0,0)(8,0) \psline{->}(0,0)(0,100) \rput(9.8,0){node strength} \rput(1.3,100){frequency} \scriptsize \psline(1,-2)(1,2) \rput(1,-6){1} \psline(2,-2)(2,2) \rput(2,-6){2} \psline(3,-2)(3,2) \rput(3,-6){3} \psline(4,-2)(4,2) \rput(4,-6){4} \psline(5,-2)(5,2) \rput(5,-6){5} \psline(6,-2)(6,2) \rput(6,-6){6}  \psline(7,-2)(7,2) \rput(7,-6){7} \psline(-0.125,20)(0.125,20) \rput(-0.45,20){20} \psline(-0.125,40)(0.125,40) \rput(-0.45,40){40} \psline(-0.125,60)(0.125,60) \rput(-0.45,60){60} \psline(-0.125,80)(0.125,80) \rput(-0.45,80){80}
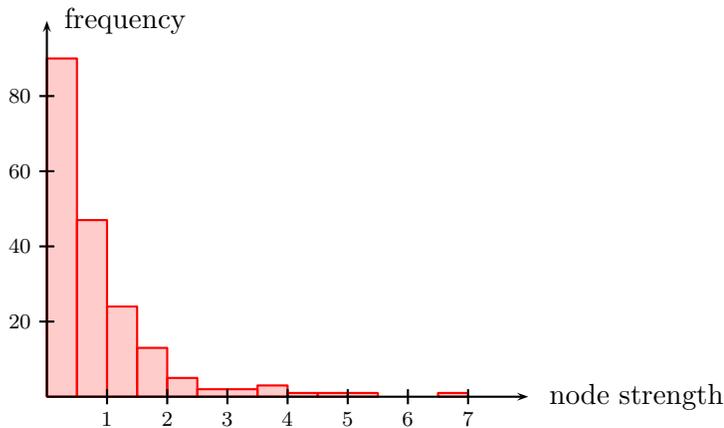
\normalsize \rput(4,-14){{\bf Figure 11.} Node strength frequency distribution.}
\end{pspicture}

\vskip 0.7 cm

\noindent \it Eigenvector centrality \rm

\vskip 0.2 cm

Node degree may be seen as a bad representation of how important a node is, for it doesn't take into account how important the neighbors of a node may be. As an example, one node may have a low dregree, but it may be connected with other nodes with very high degree, so it is, in some way, influent. A measure that takes into account the degree of neighbouring nodes when calculating the importance of a node is called eigenvector centrality. In order to define it properly, one must first define an {\sl adjency matrix}, $A$, whose elements $a_{ij}$ are 1 if there is a connection between nodes $i$ and $j$ and zero otherwise. If one now considers the eigenvectors of the adjency matrix, and choosing its largest value, one then may define the eigenvector with largest eigenvalue by the equation
\begin{equation}
\label{eigenvec}
AX=\lambda X\ ,
\end{equation}
where $X$ is the eigenvector with the largest eigenvector $\lambda $. The eigenvector centrality of a node $i$ is then defined as the $i$th element of eigenvalue $X$:
\begin{equation}
E_c=x_i\ ,
\end{equation}
where $x_i$ is the element of $X$ in row $i$.

In the present network, the vertices with highest eigenvector centralities are BBDC4 ($E_c=0.502$), VALE5 ($E_c=0.373$), VALE3 ($E_c=0.294$), and BRAP4 ($E_c=0.257$). The eigenvector centrality frequency distribution is shown in Figure 12.

\begin{pspicture}(-2,-0.5)(3.5,5.4)
\psset{xunit=20,yunit=0.04}
\pspolygon*[linecolor=lightred](0,0)(0,110)(0.02,110)(0.02,0)
\pspolygon*[linecolor=lightred](0.02,0)(0.02,25)(0.04,25)(0.04,0)
\pspolygon*[linecolor=lightred](0.04,0)(0.04,11)(0.06,11)(0.06,0)
\pspolygon*[linecolor=lightred](0.06,0)(0.06,20)(0.08,20)(0.08,0)
\pspolygon*[linecolor=lightred](0.08,0)(0.08,1)(0.1,1)(0.1,0)
\pspolygon*[linecolor=lightred](0.1,0)(0.1,9)(0.12,9)(0.12,0)
\pspolygon*[linecolor=lightred](0.12,0)(0.12,7)(0.14,7)(0.14,0)
\pspolygon*[linecolor=lightred](0.14,0)(0.14,2)(0.16,2)(0.16,0)
\pspolygon*[linecolor=lightred](0.18,0)(0.18,1)(0.2,1)(0.2,0)
\pspolygon*[linecolor=lightred](0.24,0)(0.24,1)(0.26,1)(0.26,0)
\pspolygon*[linecolor=lightred](0.28,0)(0.28,1)(0.3,1)(0.3,0)
\pspolygon*[linecolor=lightred](0.36,0)(0.36,1)(0.38,1)(0.38,0)
\pspolygon*[linecolor=lightred](0.5,0)(0.5,1)(0.52,1)(0.52,0)
\pspolygon[linecolor=red](0,0)(0,110)(0.02,110)(0.02,0)
\pspolygon[linecolor=red](0.02,0)(0.02,25)(0.04,25)(0.04,0)
\pspolygon[linecolor=red](0.04,0)(0.04,11)(0.06,11)(0.06,0)
\pspolygon[linecolor=red](0.06,0)(0.06,20)(0.08,20)(0.08,0)
\pspolygon[linecolor=red](0.08,0)(0.08,1)(0.1,1)(0.1,0)
\pspolygon[linecolor=red](0.1,0)(0.1,9)(0.12,9)(0.12,0)
\pspolygon[linecolor=red](0.12,0)(0.12,7)(0.14,7)(0.14,0)
\pspolygon[linecolor=red](0.14,0)(0.14,2)(0.16,2)(0.16,0)
\pspolygon[linecolor=red](0.18,0)(0.18,1)(0.2,1)(0.2,0)
\pspolygon[linecolor=red](0.24,0)(0.24,1)(0.26,1)(0.26,0)
\pspolygon[linecolor=red](0.28,0)(0.28,1)(0.3,1)(0.3,0)
\pspolygon[linecolor=red](0.36,0)(0.36,1)(0.38,1)(0.38,0)
\pspolygon[linecolor=red](0.5,0)(0.5,1)(0.52,1)(0.52,0)
\psline{->}(0,0)(0.54,0) \psline{->}(0,0)(0,120) \rput(0.64,0){eigenvector centrality} \rput(0.05,120){frequency} \scriptsize \psline(0.1,-2)(0.1,2) \rput(0.1,-6){0.1} \psline(0.2,-2)(0.2,2) \rput(0.2,-6){0.2} \psline(0.3,-2)(0.3,2) \rput(0.3,-6){0.3} \psline(0.4,-2)(0.4,2) \rput(0.4,-6){0.4} \psline(0.5,-2)(0.5,2) \rput(0.5,-6){0.5} \psline(-0.005,20)(0.005,20) \rput(-0.015,20){20} \psline(-0.005,40)(0.005,40) \rput(-0.015,40){40} \psline(-0.005,60)(0.005,60) \rput(-0.015,60){60} \psline(-0.005,80)(0.005,80) \rput(-0.015,80){80} \psline(-0.005,100)(0.005,100) \rput(-0.02,100){100}
\normalsize \rput(0.27,-16){{\bf Figure 12.} Eigenvector centrality frequency distribution.}
\end{pspicture}

\vskip 0.7 cm

\noindent \it Betwenness \rm

\vskip 0.2 cm

The betweennes centrality measures how much a node lies on the paths between other vertices. It is an important measure of how much a node is important as an intermediate between other nodes. It may be defined as
\begin{equation}
B_c^k=\sum_{i,j=1}^n\frac{n_{ij}^k}{m_{ij}}\ ,
\end{equation}
where $n_{ij}$ is the number of shortest paths (geodesic paths) between nodes $i$ and $j$ that pass through node $k$ and $m_{ij}$ is the total number of shortest paths between nodes $i$ and $j$. Our network is fully connected, so we need not worry about $m_{ij}$ being zero. The betweenness centrality frequency distribution is shown in Figure 13.

\begin{pspicture}(-3.5,-0.5)(3.5,6)
\psset{xunit=0.0008,yunit=0.03}
\pspolygon*[linecolor=lightred](0,0)(0,154)(500,154)(500,0)
\pspolygon*[linecolor=lightred](500,0)(500,14)(1000,14)(1000,0)
\pspolygon*[linecolor=lightred](1000,0)(1000,10)(1500,10)(1500,0)
\pspolygon*[linecolor=lightred](1500,0)(1500,2)(2000,2)(2000,0)
\pspolygon*[linecolor=lightred](2000,0)(2000,3)(2500,3)(2500,0)
\pspolygon*[linecolor=lightred](2500,0)(2500,2)(3000,2)(3000,0)
\pspolygon*[linecolor=lightred](4500,0)(4500,1)(5000,1)(5000,0)
\pspolygon*[linecolor=lightred](5000,0)(5000,1)(5500,1)(5500,0)
\pspolygon*[linecolor=lightred](9000,0)(9000,2)(9500,2)(9500,0)
\pspolygon*[linecolor=lightred](10000,0)(10000,1)(10500,1)(10500,0)
\pspolygon*[linecolor=lightred](12000,0)(12000,1)(12500,1)(12500,0)
\pspolygon[linecolor=red](0,0)(0,154)(500,154)(500,0)
\pspolygon[linecolor=red](500,0)(500,14)(1000,14)(1000,0)
\pspolygon[linecolor=red](1000,0)(1000,10)(1500,10)(1500,0)
\pspolygon[linecolor=red](1500,0)(1500,2)(2000,2)(2000,0)
\pspolygon[linecolor=red](2000,0)(2000,3)(2500,3)(2500,0)
\pspolygon[linecolor=red](2500,0)(2500,2)(3000,2)(3000,0)
\pspolygon[linecolor=red](4500,0)(4500,1)(5000,1)(5000,0)
\pspolygon[linecolor=red](5000,0)(5000,1)(5500,1)(5500,0)
\pspolygon[linecolor=red](9000,0)(9000,2)(9500,2)(9500,0)
\pspolygon[linecolor=red](10000,0)(10000,1)(10500,1)(10500,0)
\pspolygon[linecolor=red](12000,0)(12000,1)(12500,1)(12500,0)
%
%
\psline{->}(0,0)(13000,0) \psline{->}(0,0)(0,180) \rput(14500,0){betweenness} \rput(1300,180){frequency} \scriptsize \psline(1000,-2.5)(1000,2.5) \rput(1000,-7.5){1000} \psline(2000,-2.5)(2000,2.5) \rput(2000,-7.5){2000} \psline(3000,-2.5)(3000,2.5) \rput(3000,-7.5){3000} \psline(4000,-2.5)(4000,2.5) \rput(4000,-7.5){4000} \psline(5000,-2.5)(5000,2.5) \rput(5000,-7.5){5000} \psline(6000,-2.5)(6000,2.5) \rput(6000,-7.5){6000}  \psline(7000,-2.5)(7000,2.5) \rput(7000,-7.5){7000} \psline(8000,-2.5)(8000,2.5) \rput(8000,-7.5){8000} \psline(9000,-2.5)(9000,2.5) \rput(9000,-7.5){9000} \psline(10000,-2.5)(10000,2.5) \rput(10000,-7.5){10000} \psline(11000,-2.5)(11000,2.5) \rput(11000,-7.5){11000} \psline(12000,-2.5)(12000,2.5) \rput(12000,-7.5){12000} \psline(-125,20)(125,20) \rput(-450,20){20} \psline(-125,40)(125,40) \rput(-450,40){40} \psline(-125,60)(125,60) \rput(-450,60){60} \psline(-125,80)(125,80) \rput(-450,80){80} \psline(-125,100)(125,100) \rput(-500,100){100} \psline(-125,120)(125,120) \rput(-500,120){120} \psline(-125,140)(125,140) \rput(-500,140){140} \psline(-125,160)(125,160) \rput(-500,160){160}
\normalsize \rput(6500,-27){{\bf Figure 13.} Betweenness centrality frequency distribution.}
\end{pspicture}

\vskip 0.8 cm

The vertices with highest betweenness centrality are BBDC4 ($B_c=12010$), VALE5 ($B_c=10490$), VALE3 ($B_c=9197$), and GFSA3 ($B_c=9010$). There are 106 vertices with zero betweenness centrality, what means that no shortest path between any two vertices in the network pass through those vertices.

\vskip 0.3 cm

\noindent \it Closeness \rm

\vskip 0.2 cm

Another measure of centrality is the closeness centrality, which measures the average distance between one node and all the others. It is defined as the measure of the mean geodesic distance for a given node $i$, which is given by
\begin{equation}
\ell _i=\frac{1}{n}\sum_{j=1}^nd_{ij}\ ,
\end{equation}
where $n$ is the number of vertices and $d_{ij}$ is a geodesic (minimum path) distance from node $i$ to node $j$. This measure is small for highly connected vertices and large for distant or poorly connected ones. In order to obtain a measure that is large for highly connected nodes and small for poorly connected ones, one then defines the inverse closeness centrality of node $i$ as
\begin{equation}
C_c^i=\frac{1}{\ell _i}\ .
\end{equation}
The vertices with highest inverse closeness centrality are BBDC4 ($C_c=0.00107$), VALE5 ($C_c=0.00106$), GFSA3 ($C_c=0.00100$), and VALE3 ($C_c=0.00099$). The inverse closeness centrality frequency distribution is shown in Figure 14.

\begin{pspicture}(-2.5,-0.5)(3.5,4.4)
\psset{xunit=8,yunit=0.1}
\pspolygon*[linecolor=lightred](0.35,0)(0.35,1)(0.4,1)(0.4,0)
\pspolygon*[linecolor=lightred](0.4,0)(0.4,7)(0.45,7)(0.45,0)
\pspolygon*[linecolor=lightred](0.45,0)(0.45,16)(0.5,16)(0.5,0)
\pspolygon*[linecolor=lightred](0.5,0)(0.5,15)(0.55,15)(0.55,0)
\pspolygon*[linecolor=lightred](0.55,0)(0.55,21)(0.6,21)(0.6,0)
\pspolygon*[linecolor=lightred](0.6,0)(0.6,36)(0.65,36)(0.65,0)
\pspolygon*[linecolor=lightred](0.65,0)(0.65,25)(0.7,25)(0.7,0)
\pspolygon*[linecolor=lightred](0.7,0)(0.7,35)(0.75,35)(0.75,0)
\pspolygon*[linecolor=lightred](0.75,0)(0.75,7)(0.8,7)(0.8,0)
\pspolygon*[linecolor=lightred](0.8,0)(0.8,19)(0.85,19)(0.85,0)
\pspolygon*[linecolor=lightred](0.85,0)(0.85,4)(0.9,4)(0.9,0)
\pspolygon*[linecolor=lightred](0.9,0)(0.9,1)(0.95,1)(0.95,0)
\pspolygon*[linecolor=lightred](0.95,0)(0.95,1)(1,1)(1,0)
\pspolygon*[linecolor=lightred](1,0)(1,2)(1.05,2)(1.05,0)
\pspolygon[linecolor=red](0.35,0)(0.35,1)(0.4,1)(0.4,0)
\pspolygon[linecolor=red](0.4,0)(0.4,7)(0.45,7)(0.45,0)
\pspolygon[linecolor=red](0.45,0)(0.45,16)(0.5,16)(0.5,0)
\pspolygon[linecolor=red](0.5,0)(0.5,15)(0.55,15)(0.55,0)
\pspolygon[linecolor=red](0.55,0)(0.55,21)(0.6,21)(0.6,0)
\pspolygon[linecolor=red](0.6,0)(0.6,36)(0.65,36)(0.65,0)
\pspolygon[linecolor=red](0.65,0)(0.65,25)(0.7,25)(0.7,0)
\pspolygon[linecolor=red](0.7,0)(0.7,35)(0.75,35)(0.75,0)
\pspolygon[linecolor=red](0.75,0)(0.75,7)(0.8,7)(0.8,0)
\pspolygon[linecolor=red](0.8,0)(0.8,19)(0.85,19)(0.85,0)
\pspolygon[linecolor=red](0.85,0)(0.85,4)(0.9,4)(0.9,0)
\pspolygon[linecolor=red](0.9,0)(0.9,1)(0.95,1)(0.95,0)
\pspolygon[linecolor=red](0.95,0)(0.95,1)(1,1)(1,0)
\pspolygon[linecolor=red](1,0)(1,2)(1.05,2)(1.05,0)
%
%
\psline{->}(0,0)(1.2,0) \psline{->}(0,0)(0,40) \rput(1.49,0){inverse closeness ($\times 10^{-3}$)} \rput(0.13,40){frequency} \scriptsize \psline(0.1,-1)(0.1,1) \rput(0.1,-3){0.1} \psline(0.2,-1)(0.2,1) \rput(0.2,-3){0.2} \psline(0.3,-1)(0.3,1) \rput(0.3,-3){0.3} \psline(0.4,-1)(0.4,1) \rput(0.4,-3){0.4} \psline(0.5,-1)(0.5,1) \rput(0.5,-3){0.5} \psline(0.6,-1)(0.6,1) \rput(0.6,-3){0.6}  \psline(0.7,-1)(0.7,1) \rput(0.7,-3){0.7} \psline(0.8,-1)(0.8,1) \rput(0.8,-3){0.8} \psline(0.9,-1)(0.9,1) \rput(0.9,-3){0.9} \psline(1,-1)(1,1) \rput(1,-3){1} \psline(1.1,-1)(1.1,1) \rput(1.1,-3){1.1} \psline(-0.0125,10)(0.0125,10) \rput(-0.045,10){10} \psline(-0.0125,20)(0.0125,20) \rput(-0.045,20){20} \psline(-0.0125,30)(0.0125,30) \rput(-0.045,30){30}
\normalsize \rput(0.6,-9){{\bf Figure 14.} Inverse closeness centrality frequency distribution.}
\end{pspicture}

\vskip 1.2 cm

\section{Cumulative distribution function}

Note that all frequency distributions, with the exception of the one for inverse closeness centrality (the same happens for the frequency distribution of the closeness centrality), are exponentialy decreasing. One say that those frequency distributions follow a power law of the type
\begin{equation}
\label{powerlaw}
p_k=ck^{-\alpha }\ ,
\end{equation}
where $p_k$ is the frequency distribution for the value $k$, and $c$ and $\alpha $ are constants. This is a characteristic of a diversity of complex systems, and it happens in the study of earthquakes, the world wide web, networks of scientific citations, of film actors, of social interactions, protein interactions, and many other topics \cite{Newman}. Networks whose centrality measures follow this type of distribution are often called scalle-free networks, and this behavior can best be visualized if one plots a graph of the cumulative frequency distribution of a centrality in terms of the centrality values, both in logarithmic representation. Figure 15 shows the logarithm of the cumulative frequency distributions as functions of the logarithm of the five types of centrality measures we used in the last section.

\begin{pspicture}(-1,-4)(7,1)
\psset{xunit=5,yunit=1.8}
\psaxes[axesstyle=frame,xylogBase=10](1.2,-3)
\psline(1.114,-2.279) (1.079,-1.978) (1.000,-1.802) (0.903,-1.677) (0.778,-1.580) (0.778,-1.501) (0.778,-1.434) (0.699,-1.376) (0.699,-1.325) (0.699,-1.279) (0.699,-1.237) (0.699,-1.200) (0.699,-1.165) (0.699,-1.133) (0.699,-1.103) (0.602,-1.075) (0.602,-1.048) (0.602,-1.023) (0.602,-1.000) (0.602,-0.978) (0.602,-0.957) (0.602,-0.936) (0.477,-0.917) (0.477,-0.899) (0.477,-0.881) (0.477,-0.864) (0.477,-0.847) (0.477,-0.832) (0.477,-0.816) (0.477,-0.802) (0.477,-0.787) (0.477,-0.774) (0.477,-0.760) (0.477,-0.747) (0.477,-0.735) (0.477,-0.722) (0.477,-0.711) (0.477,-0.699) (0.477,-0.688) (0.477,-0.677) (0.301,-0.666) (0.301,-0.656) (0.301,-0.645) (0.301,-0.635) (0.301,-0.626) (0.301,-0.616) (0.301,-0.607) (0.301,-0.598) (0.301,-0.589) (0.301,-0.580) (0.301,-0.571) (0.301,-0.563) (0.301,-0.554) (0.301,-0.546) (0.301,-0.538) (0.301,-0.531) (0.301,-0.523) (0.301,-0.515) (0.301,-0.508) (0.301,-0.501) (0.301,-0.493) (0.301,-0.486) (0.301,-0.479) (0.301,-0.473) (0.301,-0.466) (0.301,-0.459) (0.301,-0.453) (0.301,-0.446) (0.301,-0.440) (0.301,-0.434) (0.301,-0.427) (0.301,-0.421) (0.301,-0.415) (0.301,-0.410) (0.301,-0.404) (0.301,-0.398) (0.301,-0.392) (0.301,-0.387) (0.301,-0.381) (0.301,-0.376) (0.301,-0.370) (0.301,-0.365) (0.301,-0.360) (0.301,-0.354) (0.301,-0.349) (0.000,-0.344) (0.000,-0.339) (0.000,-0.334) (0.000,-0.329) (0.000,-0.325) (0.000,-0.320) (0.000,-0.315) (0.000,-0.310) (0.000,-0.306) (0.000,-0.301) (0.000,-0.296) (0.000,-0.292) (0.000,-0.288) (0.000,-0.283) (0.000,-0.279) (0.000,-0.274) (0.000,-0.270) (0.000,-0.266) (0.000,-0.262) (0.000,-0.258) (0.000,-0.253) (0.000,-0.249) (0.000,-0.245) (0.000,-0.241) (0.000,-0.237) (0.000,-0.233) (0.000,-0.230) (0.000,-0.226) (0.000,-0.222) (0.000,-0.218) (0.000,-0.214) (0.000,-0.211) (0.000,-0.207) (0.000,-0.203) (0.000,-0.200) (0.000,-0.196) (0.000,-0.192) (0.000,-0.189) (0.000,-0.185) (0.000,-0.182) (0.000,-0.178) (0.000,-0.175) (0.000,-0.172) (0.000,-0.168) (0.000,-0.165) (0.000,-0.161) (0.000,-0.158) (0.000,-0.155) (0.000,-0.152) (0.000,-0.148) (0.000,-0.145) (0.000,-0.142) (0.000,-0.139) (0.000,-0.136) (0.000,-0.133) (0.000,-0.130) (0.000,-0.126) (0.000,-0.123) (0.000,-0.120) (0.000,-0.117) (0.000,-0.114) (0.000,-0.111) (0.000,-0.108) (0.000,-0.106) (0.000,-0.103) (0.000,-0.100) (0.000,-0.097) (0.000,-0.094) (0.000,-0.091) (0.000,-0.088) (0.000,-0.086) (0.000,-0.083) (0.000,-0.080) (0.000,-0.077) (0.000,-0.075) (0.000,-0.072) (0.000,-0.069) (0.000,-0.067) (0.000,-0.064) (0.000,-0.061) (0.000,-0.059) (0.000,-0.056) (0.000,-0.053) (0.000,-0.051) (0.000,-0.048) (0.000,-0.046) (0.000,-0.043) (0.000,-0.041) (0.000,-0.038) (0.000,-0.036) (0.000,-0.033) (0.000,-0.031) (0.000,-0.028) (0.000,-0.026) (0.000,-0.023) (0.000,-0.021) (0.000,-0.019) (0.000,-0.016) (0.000,-0.014) (0.000,-0.012) (0.000,-0.009) (0.000,-0.007) (0.000,-0.005) (0.000,-0.002) (0.000,0.000)
\rput(0.6,-3.3){a) Node degree x cumulative distribution}
\end{pspicture}
\begin{pspicture}(-5.5,-4)(1,1)
\psset{xunit=3,yunit=2.3}
\psaxes[axesstyle=frame,xylogBase=10](0,0)(-1.3,-2.5)(1,0)
\psline(0.883,-2.279) (0.723,-1.978) (0.701,-1.802) (0.588,-1.677) (0.545,-1.580) (0.507,-1.501) (0.450,-1.434) (0.419,-1.376) (0.413,-1.325) (0.395,-1.279) (0.394,-1.237) (0.384,-1.200) (0.375,-1.165) (0.372,-1.133) (0.364,-1.103) (0.356,-1.075) (0.330,-1.048) (0.324,-1.023) (0.300,-1.000) (0.296,-0.978) (0.294,-0.957) (0.292,-0.936) (0.289,-0.917) (0.289,-0.899) (0.280,-0.881) (0.278,-0.864) (0.277,-0.847) (0.271,-0.832) (0.262,-0.816) (0.222,-0.802) (0.221,-0.787) (0.185,-0.774) (0.158,-0.760) (0.154,-0.747) (0.154,-0.735) (0.149,-0.722) (0.146,-0.711) (0.145,-0.699) (0.143,-0.688) (0.140,-0.677) (0.138,-0.666) (0.129,-0.656) (0.128,-0.645) (0.122,-0.635) (0.119,-0.626) (0.119,-0.616) (0.117,-0.607) (0.117,-0.598) (0.116,-0.589) (0.115,-0.580) (0.113,-0.571) (0.109,-0.563) (0.103,-0.554) (0.102,-0.546) (0.097,-0.538) (0.086,-0.531) (0.086,-0.523) (0.086,-0.515) (0.085,-0.508) (0.083,-0.501) (0.080,-0.493) (0.078,-0.486) (0.070,-0.479) (0.069,-0.473) (0.067,-0.466) (0.058,-0.459) (0.031,-0.453) (0.031,-0.446) (0.023,-0.440) (0.020,-0.434) (0.001,-0.427) (-0.003,-0.421) (-0.021,-0.415) (-0.022,-0.410) (-0.030,-0.404) (-0.033,-0.398) (-0.033,-0.392) (-0.077,-0.387) (-0.091,-0.381) (-0.113,-0.376) (-0.114,-0.370) (-0.114,-0.365) (-0.122,-0.360) (-0.130,-0.354) (-0.131,-0.349) (-0.133,-0.344) (-0.133,-0.339) (-0.133,-0.334) (-0.137,-0.329) (-0.139,-0.325) (-0.141,-0.320) (-0.142,-0.315) (-0.142,-0.310) (-0.143,-0.306) (-0.146,-0.301) (-0.148,-0.296) (-0.151,-0.292) (-0.151,-0.288) (-0.152,-0.283) (-0.152,-0.279) (-0.154,-0.274) (-0.155,-0.270) (-0.155,-0.266) (-0.155,-0.262) (-0.157,-0.258) (-0.159,-0.253) (-0.159,-0.249) (-0.159,-0.245) (-0.160,-0.241) (-0.160,-0.237) (-0.163,-0.233) (-0.164,-0.230) (-0.165,-0.226) (-0.167,-0.222) (-0.167,-0.218) (-0.167,-0.214) (-0.171,-0.211) (-0.172,-0.207) (-0.173,-0.203) (-0.173,-0.200) (-0.173,-0.196) (-0.173,-0.192) (-0.174,-0.189) (-0.175,-0.185) (-0.175,-0.182) (-0.175,-0.178) (-0.175,-0.175) (-0.176,-0.172) (-0.180,-0.168) (-0.181,-0.165) (-0.181,-0.161) (-0.184,-0.158) (-0.185,-0.155) (-0.187,-0.152) (-0.187,-0.148) (-0.191,-0.145) (-0.191,-0.142) (-0.192,-0.139) (-0.194,-0.136) (-0.195,-0.133) (-0.196,-0.130) (-0.196,-0.126) (-0.197,-0.123) (-0.200,-0.120) (-0.201,-0.117) (-0.201,-0.114) (-0.201,-0.111) (-0.202,-0.108) (-0.204,-0.106) (-0.207,-0.103) (-0.210,-0.100) (-0.212,-0.097) (-0.214,-0.094) (-0.216,-0.091) (-0.217,-0.088) (-0.221,-0.086) (-0.221,-0.083) (-0.234,-0.080) (-0.234,-0.077) (-0.236,-0.075) (-0.240,-0.072) (-0.246,-0.069) (-0.249,-0.067) (-0.251,-0.064) (-0.254,-0.061) (-0.259,-0.059) (-0.282,-0.056) (-0.283,-0.053) (-0.285,-0.051) (-0.288,-0.048) (-0.293,-0.046) (-0.295,-0.043) (-0.297,-0.041) (-0.298,-0.038) (-0.326,-0.036) (-0.364,-0.033) (-0.375,-0.031) (-0.421,-0.028) (-0.441,-0.026) (-0.443,-0.023) (-0.447,-0.021) (-0.473,-0.019) (-0.510,-0.016) (-0.550,-0.014) (-0.579,-0.012) (-0.645,-0.009) (-0.674,-0.007) (-0.764,-0.005) (-0.788,-0.002) (-1.285,0.000)
\rput(-0.15,-2.75){b) Node strength x cumulative distribution}
\end{pspicture}

\begin{pspicture}(-7,-4)(2,3.6)
\psset{xunit=2,yunit=1.8}
\psaxes[axesstyle=frame,xylogBase=10](-3,-3)
\psline(-0.299,-2.279) (-0.428,-1.978) (-0.532,-1.802) (-0.590,-1.677) (-0.730,-1.580) (-0.815,-1.501) (-0.854,-1.434) (-0.860,-1.376) (-0.860,-1.325) (-0.866,-1.279) (-0.893,-1.237) (-0.921,-1.200) (-0.921,-1.165) (-0.921,-1.133) (-0.951,-1.103) (-0.951,-1.075) (-0.951,-1.048) (-0.991,-1.023) (-0.991,-1.000) (-0.991,-0.978) (-0.991,-0.957) (-0.991,-0.936) (-0.991,-0.917) (-1.041,-0.899) (-1.114,-0.881) (-1.125,-0.864) (-1.125,-0.847) (-1.125,-0.832) (-1.125,-0.816) (-1.155,-0.802) (-1.155,-0.787) (-1.155,-0.774) (-1.155,-0.760) (-1.155,-0.747) (-1.155,-0.735) (-1.155,-0.722) (-1.155,-0.711) (-1.155,-0.699) (-1.155,-0.688) (-1.155,-0.677) (-1.155,-0.666) (-1.167,-0.656) (-1.180,-0.645) (-1.180,-0.635) (-1.237,-0.626) (-1.244,-0.616) (-1.268,-0.607) (-1.377,-0.598) (-1.377,-0.589) (-1.377,-0.580) (-1.387,-0.571) (-1.387,-0.563) (-1.387,-0.554) (-1.398,-0.546) (-1.398,-0.538) (-1.420,-0.531) (-1.420,-0.523) (-1.420,-0.515) (-1.420,-0.508) (-1.444,-0.501) (-1.456,-0.493) (-1.456,-0.486) (-1.469,-0.479) (-1.481,-0.473) (-1.481,-0.466) (-1.495,-0.459) (-1.509,-0.453) (-1.523,-0.446) (-1.523,-0.440) (-1.523,-0.434) (-1.569,-0.427) (-1.569,-0.421) (-1.602,-0.415) (-1.658,-0.410) (-1.678,-0.404) (-1.678,-0.398) (-1.678,-0.392) (-1.678,-0.387) (-1.678,-0.381) (-1.699,-0.376) (-1.721,-0.370) (-1.745,-0.365) (-1.745,-0.360) (-1.745,-0.354) (-1.770,-0.349) (-1.770,-0.344) (-1.770,-0.339) (-1.770,-0.334) (-1.796,-0.329) (-1.796,-0.325) (-1.796,-0.320) (-1.824,-0.315) (-1.824,-0.310) (-1.824,-0.306) (-1.854,-0.301) (-1.854,-0.296) (-1.886,-0.292) (-1.886,-0.288) (-1.886,-0.283) (-1.886,-0.279) (-2.000,-0.274) (-2.000,-0.270) (-2.000,-0.266) (-2.000,-0.262) (-2.046,-0.258) (-2.046,-0.253) (-2.046,-0.249) (-2.046,-0.245) (-2.046,-0.241) (-2.046,-0.237) (-2.046,-0.233) (-2.046,-0.230) (-2.097,-0.226) (-2.097,-0.222) (-2.097,-0.218) (-2.097,-0.214) (-2.097,-0.211) (-2.155,-0.207) (-2.155,-0.203) (-2.155,-0.200) (-2.222,-0.196) (-2.222,-0.192) (-2.222,-0.189) (-2.222,-0.185) (-2.222,-0.182) (-2.301,-0.178) (-2.301,-0.175) (-2.301,-0.172) (-2.301,-0.168) (-2.301,-0.165) (-2.301,-0.161) (-2.301,-0.158) (-2.398,-0.155) (-2.398,-0.152) (-2.398,-0.148) (-2.398,-0.145) (-2.398,-0.142) (-2.398,-0.139) (-2.398,-0.136) (-2.523,-0.133) (-2.699,-0.130) (-2.699,-0.126) (-2.699,-0.123) (-2.699,-0.120) (-2.699,-0.117) (-2.699,-0.114) (-2.699,-0.111) (-2.699,-0.108) (-2.699,-0.106) (-2.699,-0.103) (-2.699,-0.100) (-2.699,-0.097) (-2.699,-0.094) (-2.699,-0.091) (-2.699,-0.088) (-2.699,-0.086) (-2.699,-0.083) (-2.699,-0.080) (-3.000,-0.077) (-3.000,-0.075) (-3.000,-0.072) (-3.000,-0.069) (-3.000,-0.067) (-3.000,-0.064) (-3.000,-0.061) (-3.000,-0.059) (-3.000,-0.056) (-3.000,-0.053) (-3.000,-0.051) (-3.000,-0.048) (-3.000,-0.046) (-3.000,-0.043) (-3.000,-0.041) (-3.000,-0.038)
\rput(-1.5,-3.3){c) Eigenvector centrality x cumulative distribution}
\end{pspicture}
\begin{pspicture}(5.4,-4)(8,3.6)
\psset{xunit=3,yunit=1.8}
\psaxes[axesstyle=frame,xylogBase=10](2,0)(2,-3)(4.3,0)
\psline(4.080,-2.279) (4.021,-1.978) (3.964,-1.802) (3.955,-1.677) (3.703,-1.580) (3.677,-1.501) (3.456,-1.434) (3.433,-1.376) (3.369,-1.325) (3.367,-1.279) (3.333,-1.237) (3.297,-1.200) (3.259,-1.165) (3.168,-1.133) (3.166,-1.103) (3.161,-1.075) (3.161,-1.048) (3.111,-1.023) (3.109,-1.000) (3.105,-0.978) (3.046,-0.957) (3.045,-0.936) (3.043,-0.917) (3.041,-0.899) (2.968,-0.881) (2.968,-0.864) (2.967,-0.847) (2.872,-0.832) (2.872,-0.816) (2.872,-0.802) (2.871,-0.787) (2.871,-0.774) (2.869,-0.760) (2.748,-0.747) (2.748,-0.735) (2.748,-0.722) (2.748,-0.711) (2.747,-0.699) (2.574,-0.688) (2.574,-0.677) (2.574,-0.666) (2.574,-0.656) (2.574,-0.645) (2.574,-0.635) (2.574,-0.626) (2.574,-0.616) (2.573,-0.607) (2.573,-0.598) (2.573,-0.589) (2.573,-0.580) (2.573,-0.571) (2.573,-0.563) (2.573,-0.554) (2.573,-0.546) (2.274,-0.538) (2.274,-0.531) (2.274,-0.523) (2.274,-0.515) (2.274,-0.508) (2.274,-0.501) (2.274,-0.493) (2.274,-0.486) (2.274,-0.479) (2.274,-0.473) (2.274,-0.466) (2.274,-0.459) (2.274,-0.453) (2.274,-0.446) (2.274,-0.440) (2.274,-0.434) (2.274,-0.427) (2.274,-0.421) (2.274,-0.415) (2.274,-0.410) (2.274,-0.404) (2.274,-0.398) (2.274,-0.392) (2.274,-0.387) (2.274,-0.381) (2.274,-0.376) (2.274,-0.370) (2.274,-0.365) (2.274,-0.360) (2.274,-0.354) (2.274,-0.349) (2,-0.349)
\rput(3.15,-3.3){d) betweenness centrality x cumulative distribution}
\end{pspicture}

\begin{pspicture}(-45,-4)(-36,3.2)
\psset{xunit=13,yunit=1.8}
\psaxes[axesstyle=frame,xylogBase=10](-2.9,0)(-3.4,-2.4)(-2.9,0)
\psline(-2.968,-2.279) (-2.976,-1.978) (-2.998,-1.802) (-3.004,-1.677) (-3.039,-1.580) (-3.044,-1.501) (-3.045,-1.434) (-3.045,-1.376) (-3.047,-1.325) (-3.047,-1.279) (-3.048,-1.237) (-3.048,-1.200) (-3.048,-1.165) (-3.048,-1.133) (-3.048,-1.103) (-3.051,-1.075) (-3.051,-1.048) (-3.055,-1.023) (-3.055,-1.000) (-3.055,-0.978) (-3.055,-0.957) (-3.055,-0.936) (-3.055,-0.917) (-3.058,-0.899) (-3.061,-0.881) (-3.067,-0.864) (-3.069,-0.847) (-3.072,-0.832) (-3.072,-0.816) (-3.073,-0.802) (-3.077,-0.787) (-3.077,-0.774) (-3.078,-0.760) (-3.078,-0.747) (-3.102,-0.735) (-3.108,-0.722) (-3.108,-0.711) (-3.108,-0.699) (-3.109,-0.688) (-3.111,-0.677) (-3.111,-0.666) (-3.111,-0.656) (-3.111,-0.645) (-3.111,-0.635) (-3.113,-0.626) (-3.113,-0.616) (-3.113,-0.607) (-3.113,-0.598) (-3.113,-0.589) (-3.113,-0.580) (-3.113,-0.571) (-3.115,-0.563) (-3.115,-0.554) (-3.115,-0.546) (-3.115,-0.538) (-3.115,-0.531) (-3.117,-0.523) (-3.117,-0.515) (-3.118,-0.508) (-3.118,-0.501) (-3.118,-0.493) (-3.118,-0.486) (-3.118,-0.479) (-3.119,-0.473) (-3.122,-0.466) (-3.124,-0.459) (-3.125,-0.453) (-3.125,-0.446) (-3.125,-0.440) (-3.127,-0.434) (-3.131,-0.427) (-3.132,-0.421) (-3.132,-0.415) (-3.132,-0.410) (-3.132,-0.404) (-3.132,-0.398) (-3.132,-0.392) (-3.132,-0.387) (-3.132,-0.381) (-3.132,-0.376) (-3.132,-0.370) (-3.132,-0.365) (-3.132,-0.360) (-3.133,-0.354) (-3.133,-0.349) (-3.134,-0.344) (-3.134,-0.339) (-3.135,-0.334) (-3.136,-0.329) (-3.136,-0.325) (-3.141,-0.320) (-3.141,-0.315) (-3.141,-0.310) (-3.141,-0.306) (-3.161,-0.301) (-3.162,-0.296) (-3.162,-0.292) (-3.163,-0.288) (-3.164,-0.283) (-3.166,-0.279) (-3.168,-0.274) (-3.168,-0.270) (-3.168,-0.266) (-3.170,-0.262) (-3.170,-0.258) (-3.170,-0.253) (-3.170,-0.249) (-3.170,-0.245) (-3.170,-0.241) (-3.170,-0.237) (-3.170,-0.233) (-3.171,-0.230) (-3.171,-0.226) (-3.171,-0.222) (-3.171,-0.218) (-3.173,-0.214) (-3.173,-0.211) (-3.173,-0.207) (-3.174,-0.203) (-3.176,-0.200) (-3.176,-0.196) (-3.178,-0.192) (-3.180,-0.189) (-3.181,-0.185) (-3.181,-0.182) (-3.182,-0.178) (-3.182,-0.175) (-3.182,-0.172) (-3.185,-0.168) (-3.187,-0.165) (-3.188,-0.161) (-3.189,-0.158) (-3.189,-0.155) (-3.191,-0.152) (-3.214,-0.148) (-3.214,-0.145) (-3.215,-0.142) (-3.215,-0.139) (-3.215,-0.136) (-3.215,-0.133) (-3.216,-0.130) (-3.217,-0.126) (-3.218,-0.123) (-3.218,-0.120) (-3.218,-0.117) (-3.218,-0.114) (-3.220,-0.111) (-3.220,-0.108) (-3.220,-0.106) (-3.220,-0.103) (-3.222,-0.100) (-3.222,-0.097) (-3.223,-0.094) (-3.223,-0.091) (-3.223,-0.088) (-3.224,-0.086) (-3.225,-0.083) (-3.229,-0.080) (-3.229,-0.077) (-3.230,-0.075) (-3.231,-0.072) (-3.231,-0.069) (-3.231,-0.067) (-3.232,-0.064) (-3.233,-0.061) (-3.237,-0.059) (-3.262,-0.056) (-3.262,-0.053) (-3.262,-0.051) (-3.263,-0.048) (-3.266,-0.046) (-3.267,-0.043) (-3.267,-0.041) (-3.269,-0.038) (-3.269,-0.036) (-3.270,-0.033) (-3.270,-0.031) (-3.270,-0.028) (-3.276,-0.026) (-3.277,-0.023) (-3.278,-0.021) (-3.278,-0.019) (-3.304,-0.016) (-3.304,-0.014) (-3.306,-0.012) (-3.307,-0.009) (-3.310,-0.007) (-3.318,-0.005) (-3.346,-0.002) (-3.356,0.000)
\rput(-3.15,-2.7){e) Inverse closeness x cumulative distribution}
\rput(-2.865,0){$10^0$} \psline(-3,-0.05)(-3,0.05) \rput(-3,0.2){$10^{-3}$} \psline(-3.2,-0.05)(-3.2,0.05) \rput(-3.2,0.2){$10^{-3.2}$} \psline(-3.4,-0.05)(-3.4,0.05) \rput(-3.4,0.2){$10^{-3.4}$}
\end{pspicture}
\noindent {\begin{minipage}{7 cm}
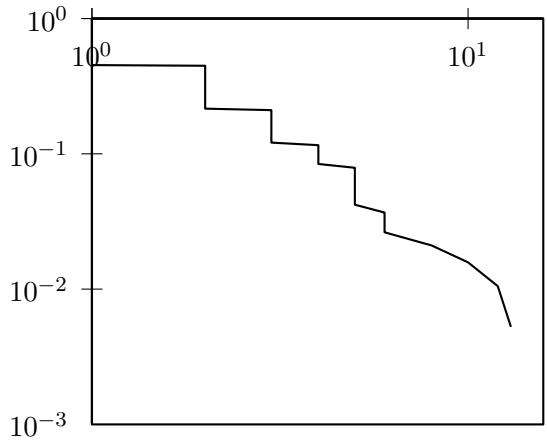
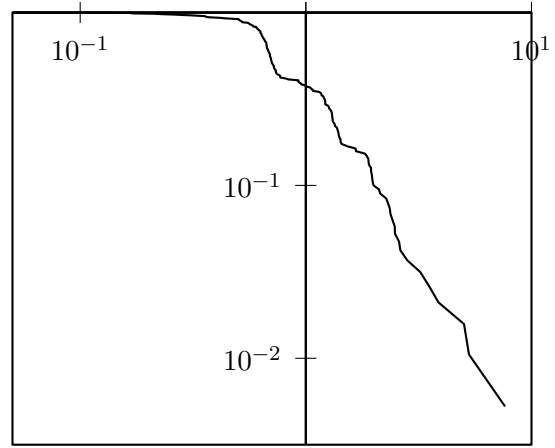
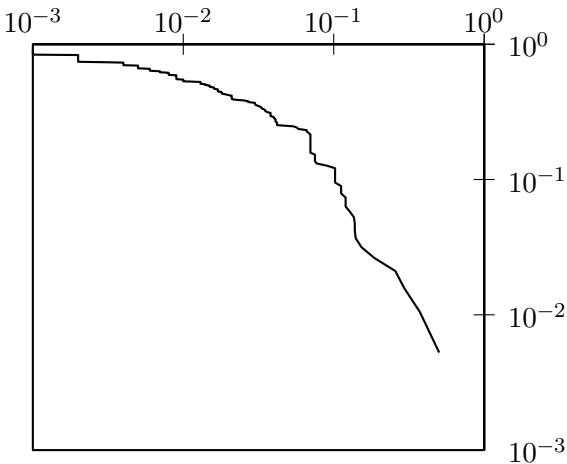
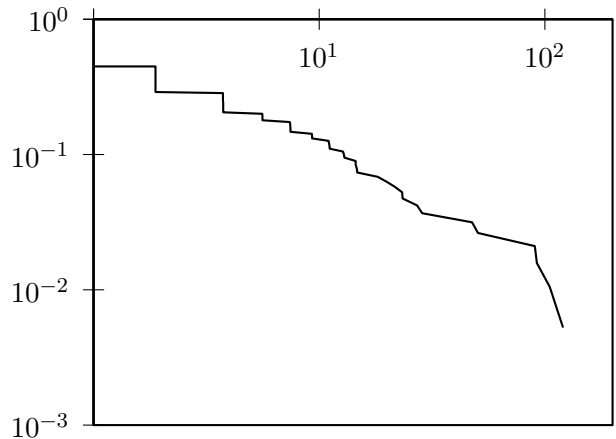
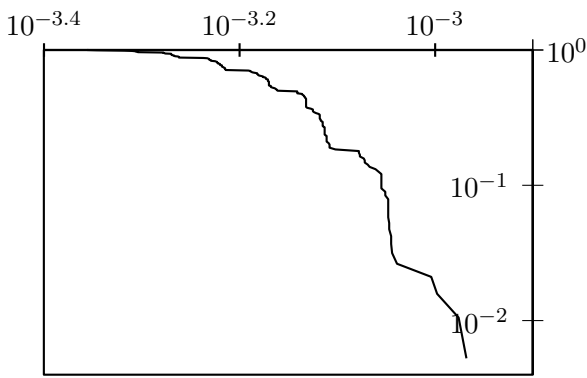
{\bf Figure 15.} Log-log plots of the cumulative frequency distributions as functions of centrality measures.
\end{minipage}

\vskip 0.9 cm

Were a network a pure scalle-free one, then the graphic of the logarithm of the cumulative frequency distribution as a function of the logarithm of a centrality measure would be a straight line. One may notice that this behavior is aproximately followed by the intermediate levels of all centrality measures, except for inverse closeness centrality. The exactly same behavior is seen in, as examples, the world wide web or networks of citations.

\section{Asset Graphs}

A two dimensional representation of non-overlapping connections between vertices, such as the minimum spanning tree, has some misleading features. Two nodes that are actualy close to each other may appear far or unconnected, and one node that is just very weakly connected may establish a connection in the MST at random. A three dimensional representation is often a better representation, and one can be built using distance as a guide. As an example, one may use an algorithm that minimizes the sum of the squares of the differences between the real distances and the ones portraied in a three dimensional graph. Other, more advanced techniques, can also be used, such as principal coordinates analysis. By using the latter, one obtains the following three dimensional representation of the stocks of BM\&F-Bovespa (Figure 16).



\vskip 1.7 cm

\begin{center}
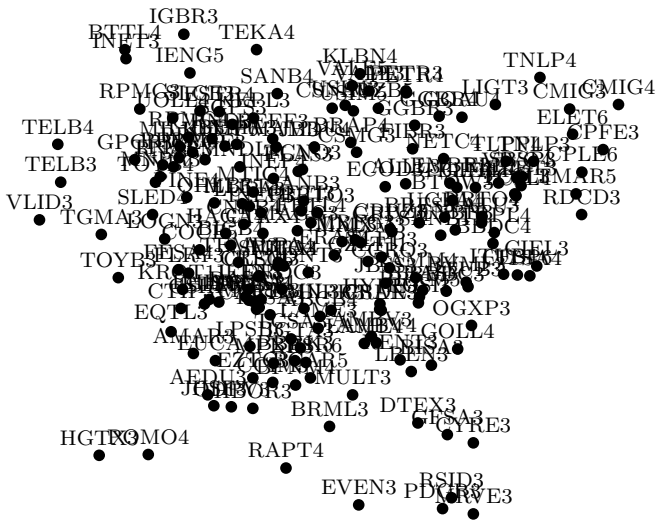
{\bf Figure 16.} Three dimensional representation of stocks of the BM\&F-Bovespa.
\end{center}

The picture is not very enlightening when printed on paper, but offers a good three-dimensional view when viewed on a computer. In particular, it may be used in order to study the distributions of stocks of the same type of companies, what is done in Appendix B. Here, we shall use the three dimensional representation in order to study asset graphs, which are networks built on the distance matrix by establishing thresholds under which connections are considered. As an example, one may build a network of those nodes whose distances with the others are below or equal to $T=0.5$. This will probably exclude many of the connections and some of the nodes of the original network. In what follows, I perform an analysis of some asset graphs obtained by selecting thresholds that go from 0.1 to 0.7, which is the limit at which random noise starts to become absolute.

At $T=0.1$ (Figure 17), the only connections are those between BBDC3 and BBDC4, both stocks from Bradesco (banking), GGBR3, GGBR4, and GOAU4 from Gerdau (metalurgy), ITUB4 and ITSA4 from Itau (banking), PETR3 and PETR4 from Petrobras (petroleum and gas), and between VALE3 and VALE5, both stocks from Vale (mining).

\begin{pspicture}(-8,-0.8)(1,2.7)
\psset{xunit=12,yunit=12,Alpha=40,Beta=0} \scriptsize
\pstThreeDLine[linecolor=black](-0.261,-0.105,-0.056)(-0.311,-0.084,-0.003) 
\pstThreeDLine[linecolor=black](-0.278,-0.180,0.128)(-0.294,-0.157,0.138) 
\pstThreeDLine[linecolor=black](-0.278,-0.180,0.128)(-0.312,-0.139,0.137) 
\pstThreeDLine[linecolor=black](-0.294,-0.157,0.138)(-0.312,-0.139,0.137) 
\pstThreeDLine[linecolor=black](-0.323,-0.080,-0.037)(-0.324,-0.036,-0.039) 
\pstThreeDLine[linecolor=black](-0.184,-0.083,0.166)(-0.188,-0.087,0.164) 
\pstThreeDLine[linecolor=black](-0.202,-0.168,0.166)(-0.187,-0.179,0.172) 
\pstThreeDDot[linecolor=blue,linewidth=1.2pt](-0.261,-0.105,-0.056) 
\pstThreeDDot[linecolor=blue,linewidth=1.2pt](-0.311,-0.084,-0.003) 
\pstThreeDDot[linecolor=blue,linewidth=1.2pt](-0.278,-0.180,0.128) 
\pstThreeDDot[linecolor=blue,linewidth=1.2pt](-0.294,-0.157,0.138) 
\pstThreeDDot[linecolor=blue,linewidth=1.2pt](-0.312,-0.139,0.137) 
\pstThreeDDot[linecolor=blue,linewidth=1.2pt](-0.324,-0.036,-0.039) 
\pstThreeDDot[linecolor=blue,linewidth=1.2pt](-0.323,-0.080,-0.037) 
\pstThreeDDot[linecolor=blue,linewidth=1.2pt](-0.184,-0.083,0.166) 
\pstThreeDDot[linecolor=blue,linewidth=1.2pt](-0.188,-0.087,0.164) 
\pstThreeDDot[linecolor=blue,linewidth=1.2pt](-0.202,-0.168,0.166) 
\pstThreeDDot[linecolor=blue,linewidth=1.2pt](-0.187,-0.179,0.172) 
\pstThreeDPut(-0.261,-0.105,-0.076){BBDC3}
\pstThreeDPut(-0.311,-0.084,0.017){BBDC4}
\pstThreeDPut(-0.278,-0.180,0.108){GGBR3}
\pstThreeDPut(-0.344,-0.157,0.158){GGBR4}
\pstThreeDPut(-0.372,-0.139,0.117){GOAU4}
\pstThreeDPut(-0.374,-0.036,-0.059){ITSA4}
\pstThreeDPut(-0.373,-0.080,-0.017){ITUB4}
\pstThreeDPut(-0.244,-0.083,0.206){PETR3}
\pstThreeDPut(-0.248,-0.087,0.184){PETR4}
\pstThreeDPut(-0.202,-0.168,0.190){VALE3}
\pstThreeDPut(-0.217,-0.179,0.142){VALE5}
\end{pspicture}

\begin{center}
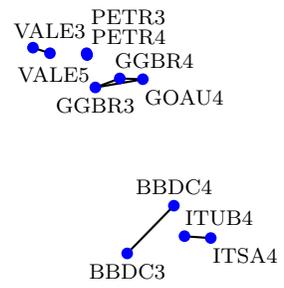
{\bf Figure 17.} Connections bellow threshold $T=0.1$.
\end{center}

\newpage

At $T=0.2$ (Figure 18), the earlier connections are joined by a small cluster of stocks RPMG3 and RPMG4, of the Refinaria de Petróleo Manguinhos (petroleum refinement), and by another small cluster formed by USIM3  and USIM4, of Usiminas (mining). Connections are established between BRAP4, Bradespar, which is an investiment branch of Bank Bradesco which has 17\% of the control of Vale, and VALE3 and VALE5. A cluster is formed with stocks from Bank Bradesco and Bank Itau, now with the joining of ITUB3.

\begin{pspicture}(-8,-1)(1,3)
\psset{xunit=12,yunit=12,Alpha=40,Beta=0} \scriptsize
\pstThreeDLine[linecolor=black](-0.261,-0.105,-0.056)(-0.311,-0.084,-0.003) 
\pstThreeDLine[linecolor=black](-0.278,-0.180,0.128)(-0.294,-0.157,0.138) 
\pstThreeDLine[linecolor=black](-0.278,-0.180,0.128)(-0.312,-0.139,0.137) 
\pstThreeDLine[linecolor=black](-0.294,-0.157,0.138)(-0.312,-0.139,0.137) 
\pstThreeDLine[linecolor=black](-0.323,-0.080,-0.037)(-0.324,-0.036,-0.039) 
\pstThreeDLine[linecolor=black](-0.184,-0.083,0.166)(-0.188,-0.087,0.164) 
\pstThreeDLine[linecolor=black](-0.202,-0.168,0.166)(-0.187,-0.179,0.172) 
\pstThreeDLine[linecolor=black](-0.324,-0.036,-0.039)(-0.311,-0.084,-0.003) 
\pstThreeDLine[linecolor=black](-0.323,-0.080,-0.037)(-0.261,-0.105,-0.056) 
\pstThreeDLine[linecolor=black](-0.323,-0.080,-0.037)(-0.311,-0.084,-0.003) 
\pstThreeDLine[linecolor=black](-0.202,-0.168,0.166)(-0.157,-0.146,0.109) 
\pstThreeDLine[linecolor=black](-0.187,-0.179,0.172)(-0.157,-0.146,0.109) 
\pstThreeDLine[linecolor=black](-0.284,-0.095,-0.051)(-0.323,-0.080,-0.037) 
\pstThreeDLine[linecolor=black](0.263,-0.009,0.146)(0.193,-0.008,0.115) 
\pstThreeDLine[linecolor=black](-0.181,-0.184,0.150)(-0.213,-0.208,0.144) 
\pstThreeDDot[linecolor=blue,linewidth=1.2pt](-0.261,-0.105,-0.056) 
\pstThreeDDot[linecolor=blue,linewidth=1.2pt](-0.311,-0.084,-0.003) 
\pstThreeDDot[linecolor=blue,linewidth=1.2pt](-0.157,-0.146,0.109) 
\pstThreeDDot[linecolor=blue,linewidth=1.2pt](-0.278,-0.180,0.128) 
\pstThreeDDot[linecolor=blue,linewidth=1.2pt](-0.294,-0.157,0.138) 
\pstThreeDDot[linecolor=blue,linewidth=1.2pt](-0.312,-0.139,0.137) 
\pstThreeDDot[linecolor=blue,linewidth=1.2pt](-0.324,-0.036,-0.039) 
\pstThreeDDot[linecolor=blue,linewidth=1.2pt](-0.284,-0.095,-0.051) 
\pstThreeDDot[linecolor=blue,linewidth=1.2pt](-0.323,-0.080,-0.037) 
\pstThreeDDot[linecolor=blue,linewidth=1.2pt](-0.184,-0.083,0.166) 
\pstThreeDDot[linecolor=blue,linewidth=1.2pt](-0.188,-0.087,0.164) 
\pstThreeDDot[linecolor=blue,linewidth=1.2pt](0.263,-0.009,0.146) 
\pstThreeDDot[linecolor=blue,linewidth=1.2pt](0.193,-0.008,0.115) 
\pstThreeDDot[linecolor=blue,linewidth=1.2pt](-0.181,-0.184,0.150) 
\pstThreeDDot[linecolor=blue,linewidth=1.2pt](-0.213,-0.208,0.144) 
\pstThreeDDot[linecolor=blue,linewidth=1.2pt](-0.202,-0.168,0.166) 
\pstThreeDDot[linecolor=blue,linewidth=1.2pt](-0.187,-0.179,0.172) 
\pstThreeDPut(-0.261,-0.105,-0.076){BBDC3}
\pstThreeDPut(-0.311,-0.084,0.017){BBDC4}
\pstThreeDPut(-0.157,-0.146,0.089){BRAP4}
\pstThreeDPut(-0.278,-0.180,0.108){GGBR3}
\pstThreeDPut(-0.344,-0.157,0.158){GGBR4}
\pstThreeDPut(-0.372,-0.139,0.117){GOAU4}
\pstThreeDPut(-0.374,-0.036,-0.059){ITSA4}
\pstThreeDPut(-0.264,-0.095,-0.031){ITUB3}
\pstThreeDPut(-0.373,-0.080,-0.017){ITUB4}
\pstThreeDPut(-0.244,-0.083,0.206){PETR3}
\pstThreeDPut(-0.248,-0.087,0.184){PETR4}
\pstThreeDPut(0.263,-0.009,0.166){RPMG3}
\pstThreeDPut(0.193,-0.008,0.095){RPMG4}
\pstThreeDPut(-0.121,-0.184,0.160){USIM3}
\pstThreeDPut(-0.143,-0.208,0.134){USIM5}
\pstThreeDPut(-0.202,-0.168,0.190){VALE3}
\pstThreeDPut(-0.217,-0.179,0.142){VALE5}
\end{pspicture}

\begin{center}
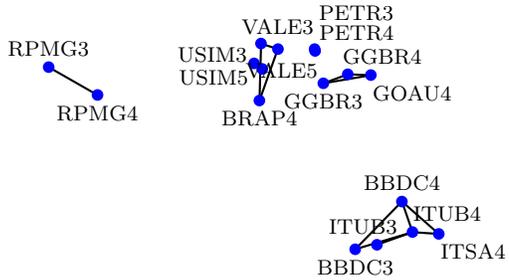
{\bf Figure 18.} Connections bellow threshold $T=0.2$.
\end{center}

At $T=0.3$ (Figure 19), the network is joined by the pairs TELB3-TELB4 of Telebras (telecommunications), INEP3-INEP4 of Inepar (construction), TNLP3-TNLP4 of Telemar (telecommunications), CMIG3-CMIG4 of CEMIG (electricity), and AMBV3-AMBV4 of Ambev (beverages). The network formed by the stocks of Gerdau now joins with the stocks of Usiminas, and with the newcommer CSNA3 of Companhia Siderúrgica Nacional (metalurgy). The new network represents a cluster of stocks of companies working in mining and meatlurgy, close to but still not connected with Vale. A new newtork is formed by the stocks GFSA3 of Gafisa, CYRE3 of Cyrela, RSID3 of Rossi Residencial, PDGR3 of PDG Realty, and MRVE3 of MRV Engenharia e Participações, all of them in the construction business.

\begin{pspicture}(-8,-3.8)(1,3)
\psset{xunit=12,yunit=12,Alpha=40,Beta=0} \scriptsize
\pstThreeDLine[linecolor=black](-0.261,-0.105,-0.056)(-0.311,-0.084,-0.003) 
\pstThreeDLine[linecolor=black](-0.278,-0.180,0.128)(-0.294,-0.157,0.138) 
\pstThreeDLine[linecolor=black](-0.278,-0.180,0.128)(-0.312,-0.139,0.137) 
\pstThreeDLine[linecolor=black](-0.294,-0.157,0.138)(-0.312,-0.139,0.137) 
\pstThreeDLine[linecolor=black](-0.284,-0.095,-0.051)(-0.324,-0.036,-0.039) 
\pstThreeDLine[linecolor=black](-0.323,-0.080,-0.037)(-0.324,-0.036,-0.039) 
\pstThreeDLine[linecolor=black](-0.184,-0.083,0.166)(-0.188,-0.087,0.164) 
\pstThreeDLine[linecolor=black](-0.202,-0.168,0.166)(-0.187,-0.179,0.172) 
\pstThreeDLine[linecolor=black](-0.324,-0.036,-0.039)(-0.311,-0.084,-0.003) 
\pstThreeDLine[linecolor=black](-0.324,-0.036,-0.039)(-0.261,-0.105,-0.056) 
\pstThreeDLine[linecolor=black](-0.284,-0.095,-0.051)(-0.261,-0.105,-0.056) 
\pstThreeDLine[linecolor=black](-0.284,-0.095,-0.051)(-0.311,-0.084,-0.003) 
\pstThreeDLine[linecolor=black](-0.323,-0.080,-0.037)(-0.261,-0.105,-0.056) 
\pstThreeDLine[linecolor=black](-0.323,-0.080,-0.037)(-0.311,-0.084,-0.003) 
\pstThreeDLine[linecolor=black](-0.202,-0.168,0.166)(-0.157,-0.146,0.109) 
\pstThreeDLine[linecolor=black](-0.187,-0.179,0.172)(-0.157,-0.146,0.109) 
\pstThreeDLine[linecolor=black](-0.284,-0.095,-0.051)(-0.323,-0.080,-0.037) 
\pstThreeDLine[linecolor=black](0.263,-0.009,0.146)(0.193,-0.008,0.115) 
\pstThreeDLine[linecolor=black](-0.181,-0.184,0.150)(-0.213,-0.208,0.144) 
\pstThreeDLine[linecolor=black](0.016,0.096,-0.106)(-0.002,0.083,-0.113) 
\pstThreeDLine[linecolor=black](-0.013,0.403,0.146)(-0.093,0.392,0.151) 
\pstThreeDLine[linecolor=black](-0.218,-0.052,-0.215)(-0.221,-0.011,-0.224) 
\pstThreeDLine[linecolor=black](-0.211,-0.242,0.147)(-0.278,-0.180,0.128) 
\pstThreeDLine[linecolor=black](-0.211,-0.242,0.147)(-0.294,-0.157,0.138) 
\pstThreeDLine[linecolor=black](-0.211,-0.242,0.147)(-0.312,-0.139,0.137) 
\pstThreeDLine[linecolor=black](-0.211,-0.242,0.147)(-0.181,-0.184,0.150) 
\pstThreeDLine[linecolor=black](-0.211,-0.242,0.147)(-0.213,-0.208,0.144) 
\pstThreeDLine[linecolor=black](-0.294,-0.157,0.138)(-0.213,-0.208,0.144) 
\pstThreeDLine[linecolor=black](-0.179,-0.014,-0.297)(-0.218,-0.052,-0.215) 
\pstThreeDLine[linecolor=black](-0.179,-0.014,-0.297)(-0.181,0.037,-0.303) 
\pstThreeDLine[linecolor=black](-0.179,-0.014,-0.297)(-0.167,0.016,-0.285) 
\pstThreeDLine[linecolor=black](-0.009,-0.124,0.031)(-0.075,-0.181,0.069) 
\pstThreeDLine[linecolor=black](-0.123,0.210,0.088)(-0.215,0.111,0.181) 
\pstThreeDLine[linecolor=black](0.332,-0.063,0.065)(0.286,-0.123,0.107) 
\pstThreeDLine[linecolor=black](-0.213,-0.208,0.144)(-0.294,-0.157,0.138) 
\pstThreeDDot[linecolor=blue,linewidth=1.2pt](0.016,0.096,-0.106) 
\pstThreeDDot[linecolor=blue,linewidth=1.2pt](-0.002,0.083,-0.113) 
\pstThreeDDot[linecolor=blue,linewidth=1.2pt](-0.261,-0.105,-0.056) 
\pstThreeDDot[linecolor=blue,linewidth=1.2pt](-0.311,-0.084,-0.003) 
\pstThreeDDot[linecolor=blue,linewidth=1.2pt](-0.157,-0.146,0.109) 
\pstThreeDDot[linecolor=blue,linewidth=1.2pt](-0.013,0.403,0.146) 
\pstThreeDDot[linecolor=blue,linewidth=1.2pt](-0.093,0.392,0.151) 
\pstThreeDDot[linecolor=blue,linewidth=1.2pt](-0.211,-0.242,0.147) 
\pstThreeDDot[linecolor=blue,linewidth=1.2pt](-0.221,-0.011,-0.224) 
\pstThreeDDot[linecolor=blue,linewidth=1.2pt](-0.218,-0.052,-0.215) 
\pstThreeDDot[linecolor=blue,linewidth=1.2pt](-0.278,-0.180,0.128) 
\pstThreeDDot[linecolor=blue,linewidth=1.2pt](-0.294,-0.157,0.138) 
\pstThreeDDot[linecolor=blue,linewidth=1.2pt](-0.312,-0.139,0.137) 
\pstThreeDDot[linecolor=blue,linewidth=1.2pt](-0.009,-0.124,0.031) 
\pstThreeDDot[linecolor=blue,linewidth=1.2pt](-0.075,-0.181,0.069) 
\pstThreeDDot[linecolor=blue,linewidth=1.2pt](-0.324,-0.036,-0.039) 
\pstThreeDDot[linecolor=blue,linewidth=1.2pt](-0.284,-0.095,-0.051) 
\pstThreeDDot[linecolor=blue,linewidth=1.2pt](-0.323,-0.080,-0.037) 
\pstThreeDDot[linecolor=blue,linewidth=1.2pt](-0.181,0.037,-0.303) 
\pstThreeDDot[linecolor=blue,linewidth=1.2pt](-0.179,-0.014,-0.297) 
\pstThreeDDot[linecolor=blue,linewidth=1.2pt](-0.184,-0.083,0.166) 
\pstThreeDDot[linecolor=blue,linewidth=1.2pt](-0.188,-0.087,0.164) 
\pstThreeDDot[linecolor=blue,linewidth=1.2pt](0.263,-0.009,0.146) 
\pstThreeDDot[linecolor=blue,linewidth=1.2pt](0.193,-0.008,0.115) 
\pstThreeDDot[linecolor=blue,linewidth=1.2pt](-0.167,0.016,-0.285) 
\pstThreeDDot[linecolor=blue,linewidth=1.2pt](-0.123,0.210,0.088) 
\pstThreeDDot[linecolor=blue,linewidth=1.2pt](-0.215,0.111,0.181) 
\pstThreeDDot[linecolor=blue,linewidth=1.2pt](0.332,-0.063,0.065) 
\pstThreeDDot[linecolor=blue,linewidth=1.2pt](0.286,-0.123,0.107) 
\pstThreeDDot[linecolor=blue,linewidth=1.2pt](-0.181,-0.184,0.150) 
\pstThreeDDot[linecolor=blue,linewidth=1.2pt](-0.213,-0.208,0.144) 
\pstThreeDDot[linecolor=blue,linewidth=1.2pt](-0.202,-0.168,0.166) 
\pstThreeDDot[linecolor=blue,linewidth=1.2pt](-0.187,-0.179,0.172) 
\pstThreeDPut(0.024,0.096,-0.086){AMBV3}
\pstThreeDPut(-0.002,0.083,-0.133){AMBV4}
\pstThreeDPut(-0.261,-0.105,-0.076){BBDC3}
\pstThreeDPut(-0.311,-0.084,0.017){BBDC4}
\pstThreeDPut(-0.157,-0.146,0.089){BRAP4}
\pstThreeDPut(-0.013,0.403,0.166){CMIG3}
\pstThreeDPut(-0.133,0.392,0.171){CMIG4}
\pstThreeDPut(-0.211,-0.242,0.167){CSNA3}
\pstThreeDPut(-0.301,-0.011,-0.224){CYRE3}
\pstThreeDPut(-0.218,-0.052,-0.195){GFSA3}
\pstThreeDPut(-0.278,-0.180,0.108){GGBR3}
\pstThreeDPut(-0.344,-0.157,0.158){GGBR4}
\pstThreeDPut(-0.372,-0.139,0.117){GOAU4}
\pstThreeDPut(-0.009,-0.124,0.011){INEP3}
\pstThreeDPut(-0.075,-0.181,0.089){INEP4}
\pstThreeDPut(-0.374,-0.036,-0.059){ITSA4}
\pstThreeDPut(-0.264,-0.095,-0.031){ITUB3}
\pstThreeDPut(-0.373,-0.080,-0.017){ITUB4}
\pstThreeDPut(-0.251,0.037,-0.303){MRVE3}
\pstThreeDPut(-0.179,-0.014,-0.317){PDGR3}
\pstThreeDPut(-0.244,-0.083,0.206){PETR3}
\pstThreeDPut(-0.248,-0.087,0.184){PETR4}
\pstThreeDPut(0.263,-0.009,0.166){RPMG3}
\pstThreeDPut(0.193,-0.008,0.095){RPMG4}
\pstThreeDPut(-0.167,0.016,-0.265){RSID3}
\pstThreeDPut(0.332,-0.063,0.045){TELB3}
\pstThreeDPut(0.286,-0.123,0.127){TELB4}
\pstThreeDPut(-0.123,0.210,0.068){TNLP3}
\pstThreeDPut(-0.215,0.111,0.201){TNLP4}
\pstThreeDPut(-0.121,-0.184,0.160){USIM3}
\pstThreeDPut(-0.143,-0.208,0.134){USIM5}
\pstThreeDPut(-0.202,-0.168,0.190){VALE3}
\pstThreeDPut(-0.217,-0.179,0.142){VALE5}
\end{pspicture}

\begin{center}
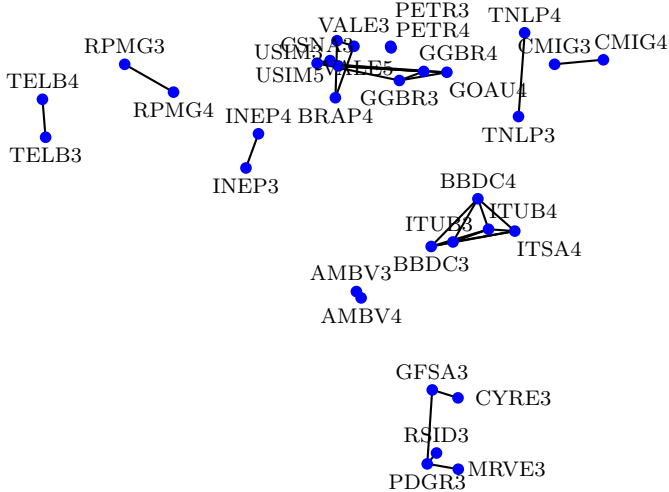
{\bf Figure 19.} Connections bellow threshold $T=0.3$.
\end{center}

For $T=0.4$ (Figure 20, with the new stocks in the network in red, for better visualization), the mining and metalurgy network joins with the financial network, and the construction network becomes denser, with more connections established between its nodes. We also have the newcommer isolated pairs TCSL3-TCSL4 of TIM (telecommunications), CIEL3 of Cielo and RDCD3 of Redecard, both operating in the business of electronic cards, SUZB5 of Suzano Papel e Celulose and FIBR3 of Fibria Celulose, both operating in the paper production market, and GOLL4 of Gol and TAMM4 of TAM, both pertaining to airlines. The stocks of two banks, SANB11 of Bank Santander and BBAS3 of the Bank of Brazil connect with the financial network via the stocks of Bradesco, and the stocks CZLT11 of Cosan (sugar and ethanol) connect with CSNA3 (metalurgy).

\begin{pspicture}(-8,-3.6)(1,3)
\psset{xunit=12,yunit=12,Alpha=40,Beta=0} \scriptsize
\pstThreeDLine[linecolor=black](-0.261,-0.105,-0.056)(-0.311,-0.084,-0.003) 
\pstThreeDLine[linecolor=black](-0.278,-0.180,0.128)(-0.294,-0.157,0.138) 
\pstThreeDLine[linecolor=black](-0.278,-0.180,0.128)(-0.312,-0.139,0.137) 
\pstThreeDLine[linecolor=black](-0.294,-0.157,0.138)(-0.312,-0.139,0.137) 
\pstThreeDLine[linecolor=black](-0.284,-0.095,-0.051)(-0.324,-0.036,-0.039) 
\pstThreeDLine[linecolor=black](-0.323,-0.080,-0.037)(-0.324,-0.036,-0.039) 
\pstThreeDLine[linecolor=black](-0.184,-0.083,0.166)(-0.188,-0.087,0.164) 
\pstThreeDLine[linecolor=black](-0.202,-0.168,0.166)(-0.187,-0.179,0.172) 
\pstThreeDLine[linecolor=black](-0.324,-0.036,-0.039)(-0.311,-0.084,-0.003) 
\pstThreeDLine[linecolor=black](-0.324,-0.036,-0.039)(-0.261,-0.105,-0.056) 
\pstThreeDLine[linecolor=black](-0.284,-0.095,-0.051)(-0.261,-0.105,-0.056) 
\pstThreeDLine[linecolor=black](-0.284,-0.095,-0.051)(-0.311,-0.084,-0.003) 
\pstThreeDLine[linecolor=black](-0.323,-0.080,-0.037)(-0.261,-0.105,-0.056) 
\pstThreeDLine[linecolor=black](-0.323,-0.080,-0.037)(-0.311,-0.084,-0.003) 
\pstThreeDLine[linecolor=black](-0.202,-0.168,0.166)(-0.157,-0.146,0.109) 
\pstThreeDLine[linecolor=black](-0.187,-0.179,0.172)(-0.157,-0.146,0.109) 
\pstThreeDLine[linecolor=black](-0.284,-0.095,-0.051)(-0.323,-0.080,-0.037) 
\pstThreeDLine[linecolor=black](0.263,-0.009,0.146)(0.193,-0.008,0.115) 
\pstThreeDLine[linecolor=black](-0.181,-0.184,0.150)(-0.213,-0.208,0.144) 
\pstThreeDLine[linecolor=black](0.016,0.096,-0.106)(-0.002,0.083,-0.113) 
\pstThreeDLine[linecolor=black](-0.013,0.403,0.146)(-0.093,0.392,0.151) 
\pstThreeDLine[linecolor=black](-0.218,-0.052,-0.215)(-0.221,-0.011,-0.224) 
\pstThreeDLine[linecolor=black](-0.211,-0.242,0.147)(-0.278,-0.180,0.128) 
\pstThreeDLine[linecolor=black](-0.211,-0.242,0.147)(-0.294,-0.157,0.138) 
\pstThreeDLine[linecolor=black](-0.211,-0.242,0.147)(-0.312,-0.139,0.137) 
\pstThreeDLine[linecolor=black](-0.211,-0.242,0.147)(-0.181,-0.184,0.150) 
\pstThreeDLine[linecolor=black](-0.211,-0.242,0.147)(-0.213,-0.208,0.144) 
\pstThreeDLine[linecolor=black](-0.211,-0.242,0.147)(-0.157,-0.146,0.109) 
\pstThreeDLine[linecolor=black](-0.211,-0.242,0.147)(-0.202,-0.168,0.166) 
\pstThreeDLine[linecolor=black](-0.211,-0.242,0.147)(-0.187,-0.179,0.172) 
\pstThreeDLine[linecolor=black](-0.211,-0.242,0.147)(-0.100,-0.048,-0.023) 
\pstThreeDLine[linecolor=black](-0.294,-0.157,0.138)(-0.213,-0.208,0.144) 
\pstThreeDLine[linecolor=black](-0.179,-0.014,-0.297)(-0.218,-0.052,-0.215) 
\pstThreeDLine[linecolor=black](-0.179,-0.014,-0.297)(-0.181,0.037,-0.303) 
\pstThreeDLine[linecolor=black](-0.179,-0.014,-0.297)(-0.167,0.016,-0.285) 
\pstThreeDLine[linecolor=black](-0.009,-0.124,0.031)(-0.075,-0.181,0.069) 
\pstThreeDLine[linecolor=black](-0.123,0.210,0.088)(-0.215,0.111,0.181) 
\pstThreeDLine[linecolor=black](0.332,-0.063,0.065)(0.286,-0.123,0.107) 
\pstThreeDLine[linecolor=black](-0.213,-0.208,0.144)(-0.294,-0.157,0.138) 
\pstThreeDLine[linecolor=black](-0.216,-0.092,-0.055)(-0.261,-0.105,-0.056) 
\pstThreeDLine[linecolor=black](-0.216,-0.092,-0.055)(-0.311,-0.084,-0.003) 
\pstThreeDLine[linecolor=black](-0.311,-0.084,-0.003)(-0.278,-0.180,0.128) 
\pstThreeDLine[linecolor=black](-0.311,-0.084,-0.003)(-0.294,-0.157,0.138) 
\pstThreeDLine[linecolor=black](-0.311,-0.084,-0.003)(-0.312,-0.139,0.137) 
\pstThreeDLine[linecolor=black](-0.311,-0.084,-0.003)(-0.226,-0.073,0.006) 
\pstThreeDLine[linecolor=black](-0.311,-0.084,-0.003)(-0.202,-0.168,0.166) 
\pstThreeDLine[linecolor=black](-0.311,-0.084,-0.003)(-0.187,-0.179,0.172) 
\pstThreeDLine[linecolor=black](-0.216,-0.092,-0.055)(-0.324,-0.036,-0.039) 
\pstThreeDLine[linecolor=black](-0.216,-0.092,-0.055)(-0.284,-0.095,-0.051) 
\pstThreeDLine[linecolor=black](-0.216,-0.092,-0.055)(-0.323,-0.080,-0.037) 
\pstThreeDLine[linecolor=black](-0.182,0.145,-0.028)(-0.230,0.165,0.029) 
\pstThreeDLine[linecolor=black](-0.221,-0.011,-0.224)(-0.181,0.037,-0.303) 
\pstThreeDLine[linecolor=black](-0.221,-0.011,-0.224)(-0.179,-0.014,-0.297) 
\pstThreeDLine[linecolor=black](-0.221,-0.011,-0.224)(-0.167,0.016,-0.285) 
\pstThreeDLine[linecolor=black](-0.188,-0.076,0.102)(-0.068,0.008,0.147) 
\pstThreeDLine[linecolor=black](-0.181,0.037,-0.303)(-0.218,-0.052,-0.215) 
\pstThreeDLine[linecolor=black](-0.181,0.037,-0.303)(-0.167,0.016,-0.285) 
\pstThreeDLine[linecolor=black](-0.312,-0.139,0.137)(-0.323,-0.080,-0.037) 
\pstThreeDLine[linecolor=black](-0.218,-0.052,-0.215)(-0.167,0.016,-0.285) 
\pstThreeDLine[linecolor=black](-0.278,-0.180,0.128)(-0.181,-0.184,0.150) 
\pstThreeDLine[linecolor=black](-0.278,-0.180,0.128)(-0.213,-0.208,0.144) 
\pstThreeDLine[linecolor=black](-0.278,-0.180,0.128)(-0.202,-0.168,0.166) 
\pstThreeDLine[linecolor=black](-0.278,-0.180,0.128)(-0.187,-0.179,0.172) 
\pstThreeDLine[linecolor=black](-0.294,-0.157,0.138)(-0.181,-0.184,0.150) 
\pstThreeDLine[linecolor=black](-0.294,-0.157,0.138)(-0.202,-0.168,0.166) 
\pstThreeDLine[linecolor=black](-0.294,-0.157,0.138)(-0.187,-0.179,0.172) 
\pstThreeDLine[linecolor=black](-0.312,-0.139,0.137)(-0.181,-0.184,0.150) 
\pstThreeDLine[linecolor=black](-0.312,-0.139,0.137)(-0.213,-0.208,0.144) 
\pstThreeDLine[linecolor=black](-0.312,-0.139,0.137)(-0.202,-0.168,0.166) 
\pstThreeDLine[linecolor=black](-0.312,-0.139,0.137)(-0.187,-0.179,0.172) 
\pstThreeDLine[linecolor=black](-0.160,0.035,-0.120)(-0.145,-0.020,-0.041) 
\pstThreeDLine[linecolor=black](-0.323,-0.080,-0.037)(-0.226,-0.073,0.006) 
\pstThreeDLine[linecolor=black](-0.323,-0.080,-0.037)(-0.202,-0.168,0.166) 
\pstThreeDLine[linecolor=black](-0.323,-0.080,-0.037)(-0.187,-0.179,0.172) 
\pstThreeDLine[linecolor=black](-0.235,0.044,0.052)(-0.237,0.044,0.050) 
\pstThreeDDot[linecolor=blue,linewidth=1.2pt](0.016,0.096,-0.106) 
\pstThreeDDot[linecolor=blue,linewidth=1.2pt](-0.002,0.083,-0.113) 
\pstThreeDDot[linecolor=blue,linewidth=1.2pt](-0.216,-0.092,-0.055) 
\pstThreeDDot[linecolor=blue,linewidth=1.2pt](-0.261,-0.105,-0.056) 
\pstThreeDDot[linecolor=blue,linewidth=1.2pt](-0.311,-0.084,-0.003) 
\pstThreeDDot[linecolor=blue,linewidth=1.2pt](-0.157,-0.146,0.109) 
\pstThreeDDot[linecolor=blue,linewidth=1.2pt](-0.182,0.145,-0.028) 
\pstThreeDDot[linecolor=blue,linewidth=1.2pt](-0.013,0.403,0.146) 
\pstThreeDDot[linecolor=blue,linewidth=1.2pt](-0.093,0.392,0.151) 
\pstThreeDDot[linecolor=blue,linewidth=1.2pt](-0.211,-0.242,0.147) 
\pstThreeDDot[linecolor=blue,linewidth=1.2pt](-0.221,-0.011,-0.224) 
\pstThreeDDot[linecolor=blue,linewidth=1.2pt](-0.100,-0.048,-0.023) 
\pstThreeDDot[linecolor=blue,linewidth=1.2pt](-0.188,-0.076,0.102) 
\pstThreeDDot[linecolor=blue,linewidth=1.2pt](-0.218,-0.052,-0.215) 
\pstThreeDDot[linecolor=blue,linewidth=1.2pt](-0.278,-0.180,0.128) 
\pstThreeDDot[linecolor=blue,linewidth=1.2pt](-0.294,-0.157,0.138) 
\pstThreeDDot[linecolor=blue,linewidth=1.2pt](-0.312,-0.139,0.137) 
\pstThreeDDot[linecolor=blue,linewidth=1.2pt](-0.160,0.035,-0.120) 
\pstThreeDDot[linecolor=blue,linewidth=1.2pt](-0.009,-0.124,0.031) 
\pstThreeDDot[linecolor=blue,linewidth=1.2pt](-0.075,-0.181,0.069) 
\pstThreeDDot[linecolor=blue,linewidth=1.2pt](-0.324,-0.036,-0.039) 
\pstThreeDDot[linecolor=blue,linewidth=1.2pt](-0.284,-0.095,-0.051) 
\pstThreeDDot[linecolor=blue,linewidth=1.2pt](-0.323,-0.080,-0.037) 
\pstThreeDDot[linecolor=blue,linewidth=1.2pt](-0.181,0.037,-0.303) 
\pstThreeDDot[linecolor=blue,linewidth=1.2pt](-0.179,-0.014,-0.297) 
\pstThreeDDot[linecolor=blue,linewidth=1.2pt](-0.184,-0.083,0.166) 
\pstThreeDDot[linecolor=blue,linewidth=1.2pt](-0.188,-0.087,0.164) 
\pstThreeDDot[linecolor=blue,linewidth=1.2pt](-0.230,0.165,0.029) 
\pstThreeDDot[linecolor=blue,linewidth=1.2pt](0.263,-0.009,0.146) 
\pstThreeDDot[linecolor=blue,linewidth=1.2pt](0.193,-0.008,0.115) 
\pstThreeDDot[linecolor=blue,linewidth=1.2pt](-0.167,0.016,-0.285) 
\pstThreeDDot[linecolor=blue,linewidth=1.2pt](-0.226,-0.073,0.006) 
\pstThreeDDot[linecolor=blue,linewidth=1.2pt](-0.068,0.008,0.147) 
\pstThreeDDot[linecolor=blue,linewidth=1.2pt](-0.145,-0.020,-0.041) 
\pstThreeDDot[linecolor=blue,linewidth=1.2pt](-0.235,0.044,0.052) 
\pstThreeDDot[linecolor=blue,linewidth=1.2pt](-0.237,0.044,0.050) 
\pstThreeDDot[linecolor=blue,linewidth=1.2pt](0.332,-0.063,0.065) 
\pstThreeDDot[linecolor=blue,linewidth=1.2pt](0.286,-0.123,0.107) 
\pstThreeDDot[linecolor=blue,linewidth=1.2pt](-0.123,0.210,0.088) 
\pstThreeDDot[linecolor=blue,linewidth=1.2pt](-0.215,0.111,0.181) 
\pstThreeDDot[linecolor=blue,linewidth=1.2pt](-0.181,-0.184,0.150) 
\pstThreeDDot[linecolor=blue,linewidth=1.2pt](-0.213,-0.208,0.144) 
\pstThreeDDot[linecolor=blue,linewidth=1.2pt](-0.202,-0.168,0.166) 
\pstThreeDDot[linecolor=blue,linewidth=1.2pt](-0.187,-0.179,0.172) 
\pstThreeDPut(0.024,0.096,-0.086){AMBV3}
\pstThreeDPut(-0.002,0.083,-0.133){AMBV4}
\pstThreeDPut(-0.176,-0.092,-0.045){\red BBAS3}
\pstThreeDPut(-0.261,-0.105,-0.076){BBDC3}
\pstThreeDPut(-0.311,-0.084,0.017){BBDC4}
\pstThreeDPut(-0.157,-0.146,0.089){BRAP4}
\pstThreeDPut(-0.182,0.145,-0.048){\red CIEL3}
\pstThreeDPut(-0.013,0.403,0.166){CMIG3}
\pstThreeDPut(-0.133,0.392,0.171){CMIG4}
\pstThreeDPut(-0.211,-0.242,0.167){CSNA3}
\pstThreeDPut(-0.301,-0.011,-0.224){CYRE3}
\pstThreeDPut(-0.030,-0.048,-0.023){\red CZLT11}
\pstThreeDPut(-0.188,-0.076,0.122){\red FIBR3}
\pstThreeDPut(-0.218,-0.052,-0.195){GFSA3}
\pstThreeDPut(-0.278,-0.180,0.108){GGBR3}
\pstThreeDPut(-0.344,-0.157,0.158){GGBR4}
\pstThreeDPut(-0.372,-0.139,0.117){GOAU4}
\pstThreeDPut(-0.160,0.035,-0.140){\red GOLL4}
\pstThreeDPut(-0.009,-0.124,0.011){INEP3}
\pstThreeDPut(-0.075,-0.181,0.089){INEP4}
\pstThreeDPut(-0.374,-0.036,-0.059){ITSA4}
\pstThreeDPut(-0.264,-0.095,-0.031){ITUB3}
\pstThreeDPut(-0.373,-0.080,-0.017){ITUB4}
\pstThreeDPut(-0.251,0.037,-0.303){MRVE3}
\pstThreeDPut(-0.179,-0.014,-0.317){PDGR3}
\pstThreeDPut(-0.244,-0.083,0.206){PETR3}
\pstThreeDPut(-0.248,-0.087,0.184){PETR4}
\pstThreeDPut(-0.230,0.165,0.049){\red RDCD3}
\pstThreeDPut(0.263,-0.009,0.166){RPMG3}
\pstThreeDPut(0.193,-0.008,0.095){RPMG4}
\pstThreeDPut(-0.167,0.016,-0.265){RSID3}
\pstThreeDPut(-0.186,-0.073,0.026){\red SANB11}
\pstThreeDPut(-0.068,0.008,0.167){\red SUZB5}
\pstThreeDPut(-0.145,-0.020,-0.021){\red TAMM4}
\pstThreeDPut(-0.235,0.044,0.072){\red TCSL3}
\pstThreeDPut(-0.237,0.044,0.030){\red TCSL4}
\pstThreeDPut(0.332,-0.063,0.045){TELB3}
\pstThreeDPut(0.286,-0.123,0.127){TELB4}
\pstThreeDPut(-0.123,0.210,0.068){TNLP3}
\pstThreeDPut(-0.215,0.111,0.201){TNLP4}
\pstThreeDPut(-0.121,-0.184,0.160){USIM3}
\pstThreeDPut(-0.143,-0.208,0.134){USIM5}
\pstThreeDPut(-0.202,-0.168,0.190){VALE3}
\pstThreeDPut(-0.217,-0.179,0.142){VALE5}
\end{pspicture}

\begin{center}
{\bf Figure 20.} Connections bellow threshold $T=0.4$.
\end{center}

For $T=0.5$ (Figure 21, with the new stocks in the network highlighted in red), all three major networks, mining and metalurgy, financial, and construction, are now joined, with a central role played by the financial network. BISA3 of Brookfield Incorporações, EVEN3 of Even Construtora e Incorporadora, EZTC3 of EZETEC, all of the building industry, join the building companies network. BRML3 of BR Malls and MULT3 of Multiplan, both of the real state business (shopping centers), form a pair close to but not connected with the building industry network.



\begin{center}
{\bf Figure 21.} Connections bellow threshold $T=0.5$.
\end{center}

BRTO4 of Brasil Telecom and TMAR5 of Telemar Norte Leste, both of telecommunications companies, join TNLP4 of Telemar in order to form a telecommunications network. GETI3 and GETI4, both of AES Tietê (electricity) form an isolated pair. LAME3 and LAME4 of Lojas Americanas and LREN3 of Lojas Renner, both of the retailing business, form another separate cluster. MMXM3 of MMX Mineração e Metálicos join Vale and Bradespar in the mining and metalurgy network. The pair of stocks of Inepar (construction) connects with Vale (mining), the pair of Telemar connects with Telemar Norte Leste, the stocks of Gol (airliner) connect with the ones of Bradesco (banking) and Gerdau (metalurgy). BVMF3, which are stocks of the BM\&F-Bovespa (financial), connect with the ones of Bradesco, Bradespar, Itau (all in the banking and financial sectors), Cyrella (building), and Gerdau (metalurgy). Petrobras (petroleum and gas) connects with Bradesco (banking) and Gerdau (metalurgy).

\newpage

For $T=0.6$ (Figure 22, with the new stocks in the network highlighted in red), and $T=0.7$, connections between sectors become more frequent, and already existing clusters become denser. More stocks take part of the complete network now, even those that have weaker correlations with other stocks. From $T=0.8$ onwards, noise starts to takes over, and new connections are not reliable.



\begin{center}
{\bf Figure 22.} Connections bellow threshold $T=0.5$.
\end{center}

\section{Centrality measures for the asset graph}

I shall now analize some of the centrality measures seen in sections 3 and 4 but now applying them to the network obtained by fixing a threshold $T=0.7$ (just bellow the noisy region) and considering only those distances that are bellow it. I begin by analyzing the node degree. Table 2 and Figure 23 show the node degree distribution.

\begin{minipage}{5 cm }
\[ \begin{array}{|c|c|} \hline \text{Node degree} & \text{Frequency} \\ \hline 0 & 22 \\ 10 & 47 \\ 20 & 22 \\ 30 & 10 \\ 40 & 14 \\ 50 & 13 \\ 60 & 16 \\ 70 & 8 \\ 80 & 9 \\ 90 & 6 \\ 100 & 4 \\ 110 & 3 \\ 120 & 6 \\ 130 & 3 \\ 140 & 2 \\ 150 & 3 \\ 160 & 2 \\ \hline \end{array} \]
{\bf Table 2.} Node degree distribution for the asset graph.
\end{minipage}
%

\vskip 0.5 cm

The highest node degrees belong to BBDC4 and ITUB4 (node degree 103), BRAP4 (node degree 101), ITSA4 (node degree 95), BBDC3, and VALE5 (node degree 95). Figure 24 represents the acumulated frequency distribution of the node degree plotted against the node degree.

%

\vskip 0.3 cm

Next, calculating the node strenght, one can see that the nodes with highest values are BBDC4 ($N_s=86.87$), ITUB4 ($N_s=85.05$), BRAP4 ($N_s=81.22$), and ITSA4 ($N_s=79.93$). The probability distribution function is shown in Figure 25, together with the log-log plot of the cumulative distribution as a function of the node strength (Figure 26).

%

\vskip 0.6 cm

The eigenvector centrality, given by equation (\ref{eigenvec}), measures the centrality of a node by giving weights to the nodes that are immediately connected with it. The highest values for eigenvector centrality are those of BBDC4 and ITUB4 ($E_c=0.172$), ITSA4 ($E_c=0.167$), BBDC3 and BRAP4 ($E_c=0.163$), and VALE5 ($E_c=0.160$). The probability distribution function is shown in figure 25, together with the log-log plot of the cumulative distribution as a function of the eigenvector centrality (figure 26).

%

\vskip 1.4 cm

If we now consider betweenness as a measure of centrality, then the highest values occur for BRAP4 ($B_c=950$), BBDC4 ($B_c=722$), ITUB4 ($B_c=701$), and VALE5 ($B_c=629$). The probability distribution function is shown in Figure 29, together with the log-log plot of the cumulative distribution as a function of the betweenness centrality (Figure 30).

\vskip 0.2 cm

%

\vskip 1.7 cm

The last centrality measure I shall analize is inverse closeness centrality, for which the highest values occur for BBDC4 and ITUB4 ($C_c=1.866\times 10^{-4}$), BRAP4 and ITSA4 ($C_c=1.864\times 10^{-4}$), BBDC3 ($C_c=1.863\times 10^{-4}$), and VALE5 ($C_c=1.862\times 10^{-4}$). The probability distribution function is shown in Figure 31, together with the log-log plot of the cumulative distribution as a function of the inverse closeness centrality (Figure 32).

%

\vskip 2.1 cm

One thing to be noticed is that the frequency distributions of the centrality measures of the asset graph are more distant from what would be expected from a scale-free network. This may be due to the great amonut of noise that comes with the choice of threshold. For higher threshold values, this difference increases, and for lower threshold values, it decreases.

\section{K-shell decomposition}

Another important centrality measure which can be applied only to the asset graph (and not to the MST) is k-shell decomposition, which consists on classifying a vertice according to the connections it makes, and also considers if it is in a region of the network that is also highly connected. This decomposition is frequently used in the study of the propagation of diseases and of information, and also shed some light on financial networks. It consists of considering all nodes with degree 1, assigning to them $k=1$, and striping the network from them. Then, one looks at the remaining vertices with degree 2 or less, and assigns to them $k=2$. Repeating the procedure, all vertices with $k=2$ are removed, and one then looks for vertices with degree 3 or less. The process goes on until all nodes are removed. What one then obtains are shells of vertices that increase in importance as $k$ goes larger. For our asset graph network for $T=0.7$, there are 30 shells, and the stocks that belong to it are represented in Figure 33.



\vskip 1.7 cm

\begin{center}
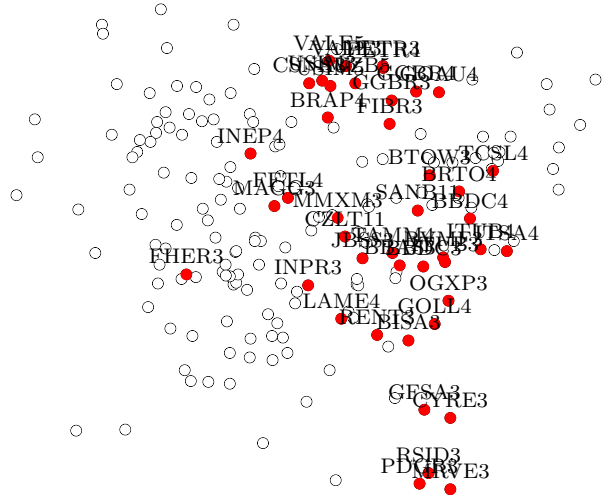
{\bf Figure 33.} Core ($k=30$) of the asset graph.
\end{center}

There is a strong dependence of k-shell position and node degree, as can be seen in Figure 34. With the exception of the innermost shell, there is an almost linear relation between the two measures.

%

\begin{center}
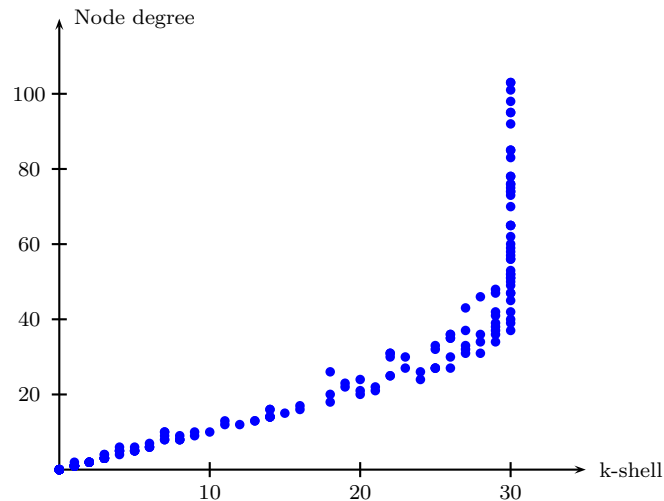
{\bf Figure 34.} Node degree as a function of $k$-shell number.
\end{center}

\section{Conclusion}

Using the concepts of minimum spanning trees and asset graphs, both based on the correlation matrix of log-returns of the BM\&F-Bovespa for the year 2010, maps of that stock exchange were built, and the network structure so obtained was examined. One could see that that the BM\&F-Bovespa has a clustered structure roughly based on economic activity sectors, and gravitates around some key stocks. The results may be used in order to obtain a better knowledge of the structure of this particular stock exchange, and maybe devise diversification strategies for portfolios of stocks belonging to it. There was also the opportunity to compare two different representations of the same complex system, each one with its benefits and maladies.

\vskip 0.6 cm

\noindent{\Large \bf Acknowledgements}

\vskip 0.4 cm

I thank for the support of this work by a grant from Insper, Instituto de Ensino e Pesquisa. I am also grateful to Leonidas Sandoval Neto (my father and Economist), who helped me clarify some details, and to Gustavo Curi Amarante, who collected the data. This article was written using \LaTeX, all figures were made using PSTricks, and the calculations were made using Matlab, Ucinet and Excel. All data are freely available upon request on leonidassj@insper.edu.br.

\appendix

\section{Stocks, codes, and sectors}

Here I display, in alphabetical order, the stocks that are being used in the present work, together with the companies they represent and the sectors those belong to. They are not all the stocks that are negotiated in the BM\&F-Bovespa, for only the ones that were negotiated every day the stock market opened are being considered. So, all the stocks have high liquidity and there is no missing data in the time series being used.

\vskip 0.3 cm

\small

\[ 
\begin{array}{|c|c|l|} \hline
\text{Code} & \text{ Company } & \text{ Sector } \\ \hline
ABCB4 & \text{ Banco ABC } & \text{ Banking } \\
AEDU3 & \text{ Anhanguera Educacional } & \text{ Education } \\
ALLL3 & \text{ América Latina Logística } & \text{ Logistics and transportation - railways } \\
ALPA4 & \text{ Alpargatas } & \text{ Consumer goods - clothing and footwear } \\
AMAR3 & \text{ Lojas Marisa } & \text{ Consumer goods - clothing and footwear } \\
AMBV3 & \text{ AMBEV } & \text{ Consumer goods - beverages } \\
AMBV4 & \text{ AMBEV } & \text{ Consumer goods - beverages } \\
\hline \end{array} \]

\[ \begin{array}{|c|c|l|} \hline
\text{Code} & \text{ Company } & \text{ Sector } \\ \hline
CTAX4 & \text{ Contax } & \text{ Technology, internet, and call-centers } \\
CTIP3 & \text{ CETIP } & \text{ Financial } \\
CYRE3 & \text{ Cyrela Brazil } & \text{ Construction, materials, and engineering } \\
CZLT11 & \text{ COSAN } & \text{ Consumer goods - food } \\
DASA3 & \text{ DASA - Diagnósticos da América } & \text{ Health Care } \\
DROG3 & \text{ Drogasil } & \text{ Consumer goods - pharmaceuticals } \\
DTEX3 & \text{ Duratex } & \text{ Construction, materials, and engineering } \\
\hline \end{array} \]

\[ \begin{array}{|c|c|l|} \hline
\text{Code} & \text{ Company } & \text{ Sector } \\ \hline
KLBN4 & \text{ Klabin } & \text{ Paper and celulosis } \\
KROT11 & \text{ Kroton Educacional } & \text{ Education } \\
LAME3 & \text{ Lojas Americanas } & \text{ Consumer goods - retail } \\
LAME4 & \text{ Lojas Americanas } & \text{ Consumer goods - retail } \\
LIGT3 & \text{ Light } & \text{ Electric energy } \\
LLXL3 & \text{ LLX Logística } & \text{ Logistics and transportation } \\
LOGN3 & \text{ Log-in Logística Intermodal } & \text{ Logistics and transportation } \\
\hline \end{array} \]

\[ \begin{array}{|c|c|l|} \hline
\text{Code} & \text{ Company } & \text{ Sector } \\ \hline
TAMM4 & \text{ TAM } & \text{ Logistics and transportation - airline } \\
TBLE3 & \text{ Tractebel } & \text{ Electric energy } \\
TCNO3 & \text{ Tecnosolo Engenharia } & \text{ Construction, materials, and engineering } \\
TCSA3 & \text{ Tecnisa } & \text{ Construction, materials, and engineering } \\
TCSL3 & \text{ TIM } & \text{ Mobile telecommunications } \\
TCSL4 & \text{ TIM } & \text{ Mobile telecommunications } \\
TEKA4 & \text{ TEKA } & \text{ Tecelagem } \\
\hline
\end{array} \]

\normalsize

\begin{center}
{\bf Table 3.} Stocks, companies, and sectors to which they belong.
\end{center}

\section{Sector distribution}

In the following figures, I present the distribution of stocks according to sectors in three dimensional maps. One may see that stocks from the same sectors tend to aglommerate.

%

\vskip 1.8 cm

\begin{center}
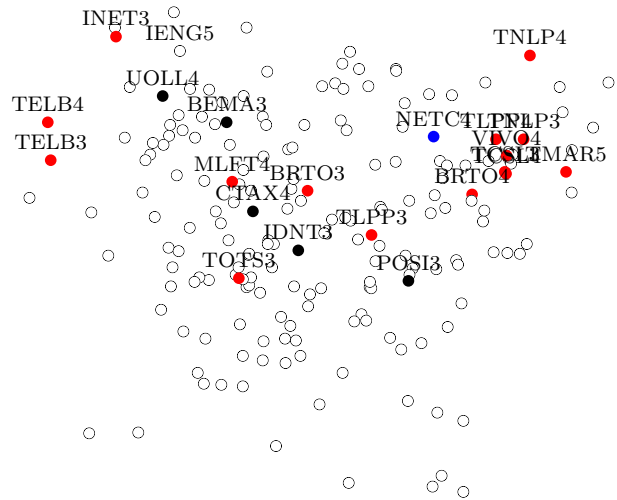
{\bf Figure B4.} Telecommunications (red), cable TV (blue), and technology, internet, and call-centers (black) sectors.
\end{center}

%

\vskip 1.8 cm

\begin{center}
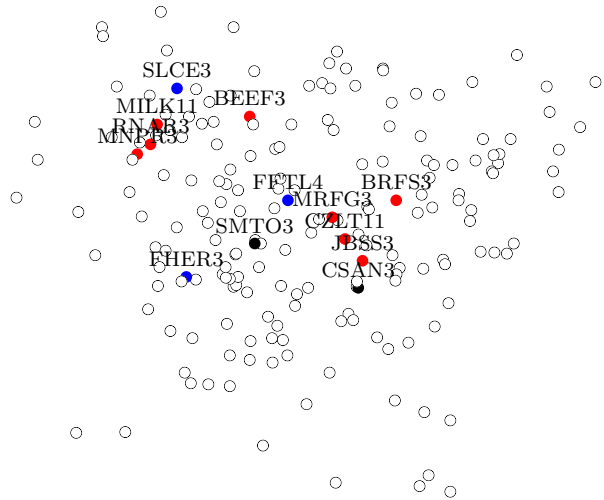
{\bf Figure B5.} Food (red), sugar and ethanol (blue), and agribusiness (black) sectors.
\end{center}

%

\vskip 1.8 cm

\begin{center}
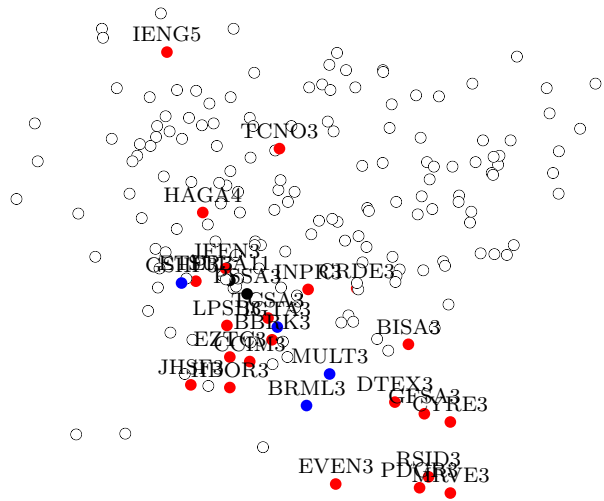
{\bf Figure B6.} Construction, materials, and engineering (red), real state (blue), and insurance (black).
\end{center}

%

\vskip 1.8 cm

\begin{center}
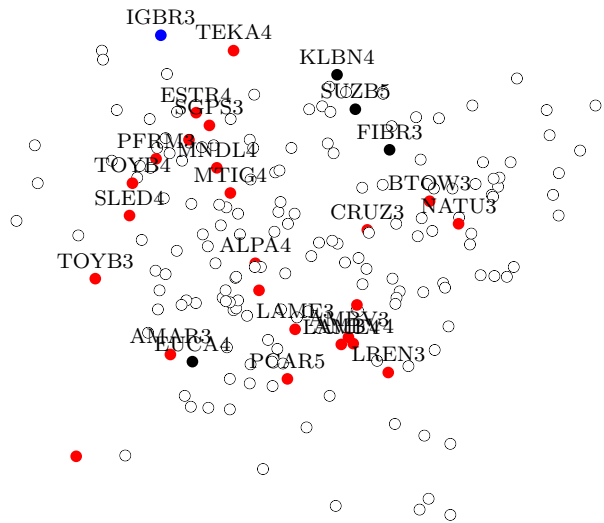
{\bf Figure B7.} Consumer goods (red), electronics (bue), and wood and paper (black).
\end{center}

%

\vskip 1.8 cm

\begin{center}
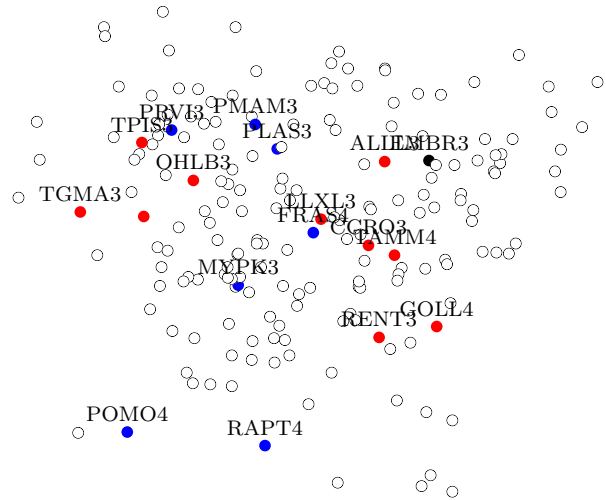
{\bf Figure B8.} Logistics and transportation (red), materials for transport and general (blue), airplane industry (black).
\end{center}

%

\vskip 1.8 cm

\begin{center}
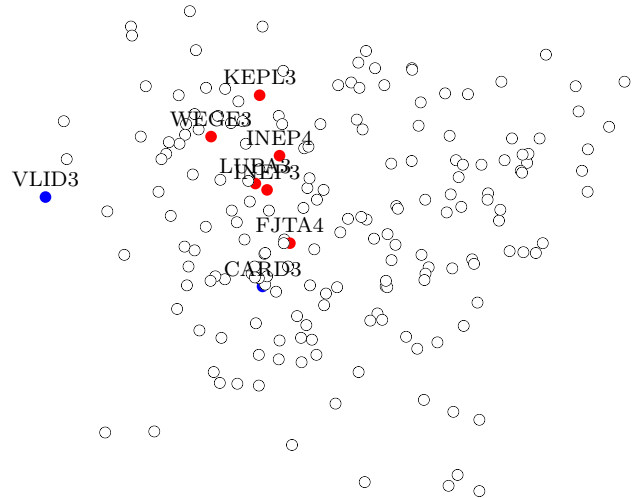
{\bf Figure B9.} Equipments and machinery (red), general services for payment (blue).
\end{center}









\end{document}